\documentclass[twocolumn]{aastex631}
\usepackage[T1]{fontenc}
\usepackage{graphicx}
\usepackage{changepage}
\let\tablenum\relax
\usepackage{siunitx}
\usepackage{amsmath}

\newcommand\ntargets{22}
\newcommand\EDENVolume{15 pc}
\newcommand\avgtime{25 hours}

\submitjournal{ApJ}

\shorttitle{Project EDEN Survey}
\shortauthors{EDEN Collaboration et al.}
\graphicspath{{./}{figures/}}

\begin{document}

\title{EDEN Survey: Small Transiting Planet Detection Limits and Constraints on the Occurrence Rates for Late M~Dwarfs within \EDENVolume{}}

\correspondingauthor{Jeremy Dietrich}
\email{jdietrich1@arizona.edu}

\author[0000-0001-6320-7410]{Jeremy Dietrich}
\affiliation{Steward Observatory and Department of Astronomy, The University of Arizona, Tucson, AZ 85721, USA}

\author[0000-0003-3714-5855]{D\'aniel Apai}
\affiliation{Steward Observatory and Department of Astronomy, The University of Arizona, Tucson, AZ 85721, USA}
\affiliation{Lunar and Planetary Laboratory and Department of Planetary Sciences, The University of Arizona, Tucson, AZ 85721, USA}

\author[0000-0001-8355-2107]{Martin Schlecker}
\affiliation{Steward Observatory and Department of Astronomy, The University of Arizona, Tucson, AZ 85721, USA}
\affiliation{Max-Planck-Institut f\"ur Astronomie, K\"onigstuhl 17, 69117 Heidelberg, Germany}

\author[0000-0003-3702-0382]{Kevin K. Hardegree-Ullman}
\affiliation{Steward Observatory and Department of Astronomy, The University of Arizona, Tucson, AZ 85721, USA}

\author[0000-0002-3627-1676]{Benjamin V. Rackham}
\altaffiliation{51 Pegasi b Fellow}
\affiliation{Department of Physics and Kavli Institute for Astrophysics and Space Research, Massachusetts Institute of Technology, Cambridge, MA 02139,
USA}
\affiliation{Department of Earth, Atmospheric and Planetary Science, Massachusetts Institute of Technology, 77 Massachusetts Avenue, Cambridge, MA 02139,
USA}

\author[0000-0002-2358-4796]{Nicolas Kurtovic}
\affiliation{Max-Planck-Institut f\"ur Astronomie, K\"onigstuhl 17, 69117 Heidelberg, Germany}

\author[0000-0002-0502-0428]{Karan Molaverdikhani}
\affiliation{Universit\"ats-Sternwarte, Ludwig-Maximilians-Universität
München, Scheinerstrasse 1, 81679 M\"unchen, Germany}
\affiliation{Exzellenzcluster Origins, Boltzmannstrasse 2, 85748 Garching,
Germany}
\affiliation{Max-Planck-Institut f\"ur Astronomie, K\"onigstuhl 17, 69117 Heidelberg, Germany}
\affiliation{Landessternwarte, Zentrum f\"ur Astronomie der Universität Heidelberg, K\"onigstuhl 12, 69117 Heidelberg, Germany}

\author[0000-0002-1315-9307]{Paul Gabor}
\affiliation{Vatican Observatory, 00120 Citt\`a del Vaticano}

\author{Thomas Henning}
\affiliation{Max-Planck-Institut f\"ur Astronomie, K\"onigstuhl 17, 69117 Heidelberg, Germany}

\author[0000-0003-0262-272X]{Wen-Ping Chen}
\affiliation{Graduate Institute of Astronomy, National Central University, Zhongli, Taoyuan 32001, Taiwan}

\author[0000-0002-9428-8732]{Luigi Mancini}
\affiliation{Department of Physics, University of Rome ``Tor Vergata'', Via della Ricerca Scientifica 1, I-00133, Rome, Italy}
\affiliation{Max-Planck-Institut f\"ur Astronomie, K\"onigstuhl 17, 69117 Heidelberg, Germany}
\affiliation{INAF — Turin Astrophysical Observatory, via Osservatorio 20, I-10025, Pino Torinese, Italy}
\affiliation{International Institute for Advanced Scientific Studies (IIASS), Via G. Pellegrino 19, I-84019, Vietri sul Mare (SA), Italy}

\author[0000-0003-2831-1890]{Alex Bixel}
\affiliation{Steward Observatory and Department of Astronomy, The University of Arizona, Tucson, AZ 85721, USA}

\author[0000-0002-9027-4456]{Aidan Gibbs}
\affiliation{Department of Physics \& Astronomy, University of California, Los Angeles, CA 90095, USA}

\author{Richard P. Boyle}
\affiliation{Vatican Observatory, 00120 Citt\`a del Vaticano}

\author[0000-0001-6871-4163]{Samantha Brown-Sevilla}
\affiliation{Max-Planck-Institut f\"ur Astronomie, K\"onigstuhl 17, 69117 Heidelberg, Germany}

\author[0000-0002-9020-7309]{Remo Burn}
\affiliation{Max-Planck-Institut f\"ur Astronomie, K\"onigstuhl 17, 69117 Heidelberg, Germany}

\author[0000-0002-7901-6157]{Timmy N. Delage}
\affiliation{Max-Planck-Institut f\"ur Astronomie, K\"onigstuhl 17, 69117 Heidelberg, Germany}

\author[0000-0001-8906-1528]{Lizxandra Flores-Rivera}
\affiliation{Max-Planck-Institut f\"ur Astronomie, K\"onigstuhl 17, 69117 Heidelberg, Germany}

\author[0000-0002-8889-2992]{Riccardo Franceschi}
\affiliation{Max-Planck-Institut f\"ur Astronomie, K\"onigstuhl 17, 69117 Heidelberg, Germany}

\author[0000-0003-3622-8712]{Gabriele Pichierri}
\affiliation{Max-Planck-Institut f\"ur Astronomie, K\"onigstuhl 17, 69117 Heidelberg, Germany}

\author[0000-0002-6639-1628]{Sofia Savvidou}
\affiliation{Max-Planck-Institut f\"ur Astronomie, K\"onigstuhl 17, 69117 Heidelberg, Germany}

\author[0000-0003-4322-8120]{Jonas Syed}
\affiliation{Max-Planck-Institut f\"ur Astronomie, K\"onigstuhl 17, 69117 Heidelberg, Germany}

\author[0000-0002-1560-4590]{Ivan Bruni}
\affiliation{INAF – OAS, Osservatorio di Astrofisica e Scienza dello Spazio di Bologna, Via P. Gobetti 93/3, 40129 – Bologna, Italy}

\author[0000-0002-3140-5014]{Wing-Huen Ip}
\affiliation{Graduate Institute of Astronomy, National Central University, Zhongli, Taoyuan 32001, Taiwan}

\author[0000-0001-8771-7554]{Chow-Choong Ngeow}
\affiliation{Graduate Institute of Astronomy, National Central University, Zhongli, Taoyuan 32001, Taiwan}

\author[0000-0002-3211-4219]{An-Li Tsai}
\affiliation{Graduate Institute of Astronomy, National Central University, Zhongli, Taoyuan 32001, Taiwan}

\author[0000-0001-5989-7594]{Chia-Lung Lin}
\affiliation{Graduate Institute of Astronomy, National Central University, Zhongli, Taoyuan 32001, Taiwan}

\author{Wei-Jie Hou}
\affiliation{Graduate Institute of Astronomy, National Central University, Zhongli, Taoyuan 32001, Taiwan}

\author{Hsiang-Yao Hsiao}
\affiliation{Graduate Institute of Astronomy, National Central University, Zhongli, Taoyuan 32001, Taiwan}

\author{Chi-Sheng Lin}
\affiliation{Graduate Institute of Astronomy, National Central University, Zhongli, Taoyuan 32001, Taiwan}

\author{Hung-Chin Lin}
\affiliation{Graduate Institute of Astronomy, National Central University, Zhongli, Taoyuan 32001, Taiwan}

\author[0000-0003-4508-2436]{Ritvik Basant}
\affiliation{Steward Observatory and Department of Astronomy, The University of Arizona, Tucson, AZ 85721, USA}


\collaboration{40}{Project EDEN}


\begin{abstract}
Earth-sized exoplanets that transit nearby, late spectral type red dwarfs will be prime targets for atmospheric characterization in the coming decade. Such systems, however, are difficult to find via wide-field transit surveys like \textit{Kepler} or TESS. Consequently, the presence of such transiting planets is unexplored and the occurrence rates of short-period Earth-sized planets around late M~dwarfs remain poorly constrained. Here, we present the deepest photometric monitoring campaign of \ntargets{} nearby late M~dwarf stars, using data from over 500 nights on seven 1-2 meter class telescopes. Our survey includes all known single quiescent northern late M~dwarfs within 15 pc.  We use transit-injection-and-recovery tests to quantify the completeness of our survey, successfully identify most ($>80\%$) transiting short-period (0.5-1 d) super-Earths (R $>1.9 R_\oplus$), and are sensitive ($\sim$50\%) to transiting Earth-sized planets (1.0--1.2 $R_\oplus$). Our high sensitivity to transits with a near-zero false positive rate demonstrates an efficient survey strategy. Our survey does not yield a transiting planet detection, yet it provides the most sensitive upper limits on transiting planets orbiting our target stars. Finally, we explore multiple hypotheses about the occurrence rates of short-period planets (from Earth-sized planets to giant planets) around late M~dwarfs. We show, for example, that giant planets at short periods ($<$ 1 day) are uncommon around our target stars. Our dataset provides some insight into occurrence rates of short-period planets around TRAPPIST-1-like stars, and our results can help test planetary formation and system evolution models, as well as guide future observations of nearby late M~dwarfs.

\end{abstract}



\section{Introduction} \label{sec:intro}

Due to their intrinsic faintness, the in-depth characterization (beyond transit spectroscopy) of Earth-sized exoplanets will remain, for the foreseeable future, limited to the closest of planets -- those that are within approximately \EDENVolume{} \citep[e.g.,][]{HabEx,LUVOIR,LIFE,Nautilus}. Although there are about 1000 stars in that volume \citep[][]{GCNS2021} and likely at least 1 planet per star \citep[e.g.,][]{Hardegree-Ullman2019}, as of July 2022 only 175 planets have been confirmed, mostly via radial velocity (RV) measurements.\footnote{\url{https://exoplanetarchive.ipac.caltech.edu/}} Even in this small volume of the Milky Way surrounding us, a large fraction of planets remains undiscovered. Of the planets known, a rarely found type stands out: small planets ($R_p<1.8 R_\oplus$) transiting small host stars ($R_*<0.4 R_\odot$), which are ideal targets for transmission spectroscopic studies due to the deeper transits and generally higher signal-to-noise ratio (SNR) than planets around larger host stars.

Arguably, the best-suited terrestrial planets for in-depth spectroscopic studies, both due to their number and transit observability, are currently in the TRAPPIST-1 planetary system \citep[][]{Gillon2016,Gillon2017}. These seven approximately Earth-sized planets with similar or slightly lower densities than that of Earth \citep[][]{Agol2021} are orbiting a very late spectral type M~dwarf (M8), near the stellar-substellar boundary. However, such low-mass stars are intrinsically very faint, limiting the feasibility of transit spectroscopy to systems that are very close to us (typically d $<15$~pc). Identifying a greater number of broadly TRAPPIST-1-like planetary systems would provide important opportunities for in-depth characterization of Earth-sized planets. Furthermore, a better understanding of the formation and evolution of these planets can provide context for the interpretation of incomplete data on their present-day atmospheres \citep[e.g.,][]{Apai2018}.

Yet, the occurrence rates and even the formation and evolution of these planetary systems remains poorly understood.  Important population-level constraints emerged from the analysis of planetary systems around earlier spectral type stars that can be more readily studied via wide area transit surveys like \textit{Kepler} \citep[][]{Borucki2010} and TESS \citep[][]{Ricker2015}. With a small set of detected planets around mid-M~dwarfs from \textit{Kepler} data, \citet{Hardegree-Ullman2019} showed evidence for a higher occurrence rate of short-period (0.5--10 days) small planets (0.5--2.5 $R_{\oplus}$) for mid-M~dwarfs ($1.19^{+0.70}_{-0.49}$ planets per star) than for early M~dwarfs ($0.63^{+0.08}_{-0.06}$ planets per star), suggesting that the trend continues to the lowest mass stars, consistent with the lower limit on planet occurrence around late-M dwarfs implied  by the detection of TRAPPIST-1 \citep[][]{Lienhard2020}.

\begin{figure*}[ht]
    \centering
    \includegraphics[width=2\columnwidth]{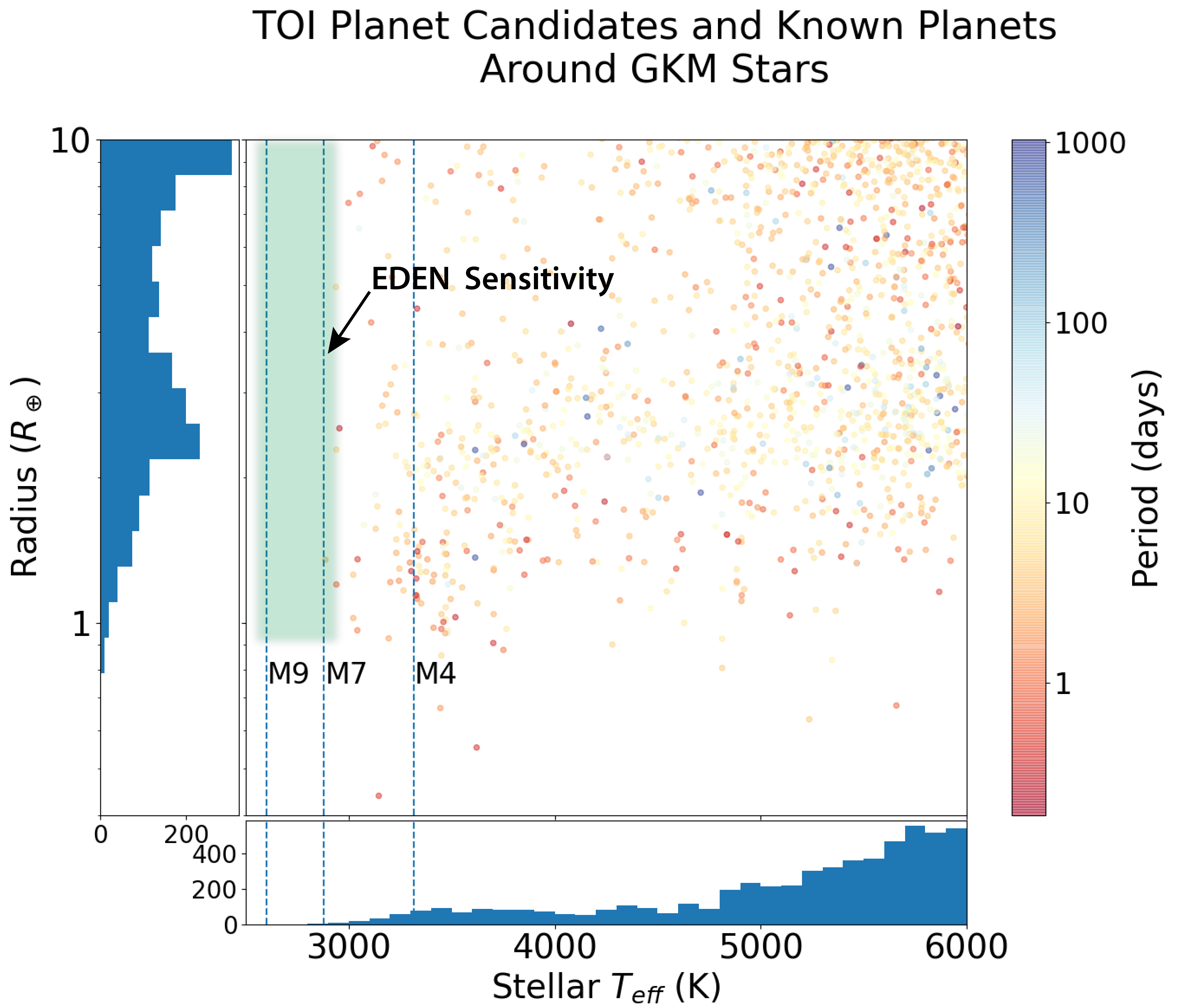}
    \caption{TESS objects of interest (TOIs) by stellar effective temperature and planet candidate radius. The temperature and planet radius distributions of the TOIs show a marked drop toward late spectral types ($>$M4--M6) and for small planets ($R_p<2$~$R_\oplus$). The green shaded region represents the sensitivity of our EDEN survey, which is complementary to that of TESS.}
    \label{fig:TESSTOIST}
\end{figure*}

However, most current surveys studying planets around M~dwarfs have some level of biases. Only 1\% of the 186000 stars in the main Kepler sample, for example, are M~dwarfs \citep[][]{Berger2020}, so studies of transiting M~dwarfs from Kepler \citep[e.g.,][]{Dressing2015,Mulders2015} are a biased sample, as these are likely only the brightest and earliest M~dwarfs at a relatively far distance. The K2 stellar catalog had a larger sample of M~dwarfs with $\sim 17\%$ of 220000 stars \citep[][]{Hardegree-Ullman2020}, but this is still underrepresented as compared to the $\sim 70\%$ of stars in the galaxy that are M~dwarfs \citep[][]{Henry2006} and again biased towards the larger, brighter M~dwarfs. K2 studies of M~dwarfs \citep[e.g.,][]{Sagear2020, SestovicDemory2020} found no planets in a targeted sample list of $\sim800-900$ stars, but this was not a blind volume-limited search, as it was guest-observer driven and therefore biased. RV surveys like CARMENES and HADES~\citep[][Ribas et al. (accepted)]{Sabotta2021,Pinamonti2022} are simply not sensitive to later M~dwarfs with the current technological limits, as is also seen with the current TESS sample (see Figure~\ref{fig:TESSTOIST}).

Therefore, targeted surveys to detect exoplanets around mid-to-late-M~dwarfs and early-L~dwarfs like EDEN, MEarth \citep[][]{Irwin2009}, SPECULOOS \citep[][]{Delrez2018,Sebastian2021} and PINES \citep[][]{Tamburo2022} are necessary to better understand planet occurrence near the substellar boundary. Even though their current detection numbers are low (less than 10 currently), these blind, volume-limited surveys still provide the least biased study of mid- to late-M~dwarfs. Improved occurrence rate estimates are an important feedback for planet formation theories, which can be constrained by comparing synthesized planet populations to observed samples \citep[e.g.,][]{Ida2004,Thommes2008,Mordasini2009b,Ndugu2018,Izidoro2021b,Schlecker2021a,Schlecker2021b,Bitsch2021,Mulders2021,Kimura2022}


The Exoearths Discovery and Exploration Network (EDEN) is a global network of professional telescopes that works toward increasing our understanding of potentially habitable exoplanetary systems targeting the detection of, in particular, planetary systems within 15~pc. EDEN is led by four Co-PIs (D. Apai, P. Gabor, Th. Henning, W-P. Chen) on behalf of four institutions (Univ. Arizona, Vatican Observatory, Max Planck Institute for Astronomy in Heidelberg, and National Central University in Taiwan). A complementary, major survey component using research telescopes in Italy was led by L. Mancini (University of Rome ``Tor Vergata’’). EDEN began its full survey with seven research telescopes (diameters between 0.8--2.3~m) surveying nearby stars in 2018 for transiting planets. EDEN's primary targets are late M~dwarf stars beyond the reach of TESS (see Figure~\ref{fig:TESSTOIST}), and for these targets EDEN demonstrated sensitivity down to Earth-sized exoplanets \citep[][]{Gibbs2020}. In this study, we observed over 500 nights on the northern late M~dwarf population within \EDENVolume{} (with at least \avgtime{} per target) looking for transiting planets. In addition to its core transit search mission, EDEN also published studies of stellar flares and their impact on the habitability of the nearby star \object{Wolf 359} \citep[][]{Lin2021}, and contributed to many follow-up observations of TESS-identified planet candidates \citep[][Peterson et al., submitted; Pozuelos et al., submitted]{Wells2021}.


This paper is organized as follows. In Section 2 we describe the Project EDEN observing strategy, with the data reduction and detrending procedures shown in Section 3.  We explain our analysis for the transit search and sensitivity in Section 4, and Section 5 contains the results of these analyses. Section 6 contains the implementation and results of our planet population hypothesis testing. Finally, in Section 7 we discuss our results from a population statistics perspective, as well as an unconfirmed signal, before summarizing our study in Section 8.


\section{Observations} \label{sec:obs}

We continued EDEN observations as laid out in \citet[][]{Gibbs2020}.  This section describes the key points of the observing strategy, including the procedure and the target stars for our survey.

\subsection{Observing Process} \label{subsec:procedure}

We observed with seven telescopes across the northern hemisphere: four located in Arizona, two in Europe, and one in Taiwan \citep[see Table 1 in][]{Gibbs2020}.  Under ideal scheduling and weather conditions, this longitudinal coverage allows for continuous observations of target stars visible from all EDEN sites. The telescope apertures range in size from 0.8~m to 2.3~m. The Schulman 0.8~m telescope is robotically operated, while the CAHA 1.2~m is remotely operated. The rest require an on-site operator/observer each night. We took sky flat or dome flats, as well as bias frames, whenever possible. Dark frames are no longer acquired as we found no difference between data calibrated with or without darks, as the dark current is near zero at all telescopes and the targets have high SNR \citep[][]{Gibbs2020}.  We tried to observe a given target star for at least three hours per night, which is roughly three times the expected transit duration of a planet on a close-in orbit. This strategy allows us to observe multiple targets in one night if we cannot spend the entire night on one star (e.g., if the star is relatively southern and is only visible for half the night).

We used the GG495 filter from Schott\footnote{\url{https://www.schott.com/shop/advanced-optics/en/Matt-Filter-Plates/GG495/c/glass-GG495}}, which is a long-pass filter that reaches 50\% transparency at $\sim$500 nm and extends into the near-infrared, due to its greater transmission ($>90$\%) than narrowband filters across visible and near-IR wavelengths. Some observations required a narrowband filter, usually due to moon brightness or for TESS transit follow-up.  We usually used 60-s exposures unless the target was too bright/faint, or if higher time precision was required (e.g., transit timing variation (TTV) analysis of transit follow-up targets).  In certain cases, we slightly defocused the star to both mitigate inter- and intra-pixel sensitivity differences and to avoid saturation for longer exposures on brighter targets \citep[for further information see, e.g.,][]{Southworth2009, Gibbs2020}.

Our detectors enter a non-linear regime of photon-electron counts at about 2/3 of the well limit; we thus aimed to stay below 60\% of the full well level for bright targets.  Our data reduction pipeline adjusts for saturation in the target and best reference stars, with brighter reference stars discarded if they are saturated.  For most of our CCDs, the field-of-view is on the order of 10\,arcmin, but the plate pixel scale differs, and in many cases the observations were binned in order to decrease readout time.  We built our data reduction pipeline to be robust and agnostic to different plate scales, and only run photometry for one given night from one given telescope at a time.

\subsection{Target Stars} \label{subsec:targets}

The Project EDEN target list is comprised of all known single non-variable stars of spectral type M7--L0 within \EDENVolume{} and with declinations > -20$^\circ$.  At the beginning of our survey, we removed known binaries and active flare stars as indicated by SIMBAD from the 15~pc sample because they are likely to contaminate the light curve, making it more difficult to find a transiting planet.  Once we removed the stars that do not fit the above criteria, we were left with \ntargets{} stars that became the focus of the survey.

It has been shown that the removal of binary stars from samples of demographic studies can introduce biases, mainly in the form of overestimating giant planet occurrence rates \citep{Moe2021}. While close binaries may pose a threat to M~dwarf planetary systems, they are relatively rare around low-mass stars \citep[e.g.,][]{Shan2015, Susemiehl2022}. M~dwarf binaries also tend to have 10$\times$ higher flare activity than single M~dwarfs \citep[][]{Huang2020}. The close binary fraction is also relatively constant across the M~dwarf range, which suggests that there is no introduction of a systematic error with their removal \citep{Offner2022}. Additionally, M~dwarf binary systems with planets or planet candidates tend to have wide separations between the stellar components \citep[e.g.,][]{Clark2022a}. To identify multiple stars in our sample, we searched the RECONS M~dwarf binary catalogs \citep{Winters2019,Vrijmoet2020}, the Gaia EDR3 binary catalog \citep{El-Badry2021}, the recent results from Robo-AO \citep{Salama2021,Salama2022}, the M~dwarf TESS objects of interest companion catalog \citep{Clark2022a}, and the POKEMON speckle catalog \citep{Clark2022b}.  We found two binaries within our sample, EDEN Input Catalog (EIC)~10 and EIC~14.

\citet{Reiners2009} identified EIC~10 (LP~775-031/Gaia~DR3~3171631420210205056) as a double-lined spectroscopic binary (SB2) consisting of two nearly equal brightness components. Recent high-resolution imaging efforts have been unable to resolve the companion \citep{Winters2019,Vrijmoet2020}. Clark et al.\ (in prep) observed EIC~10 using speckle imaging, but were not able to resolve a companion. Speckle imaging is sensitive to separations $>0\farcs1$, constraining the binary separation to $\lesssim1$~au at a system distance of 10.6~pc. Indeed, Gaia DR3 revealed this target to have an orbital astrometric solution with a period of 105.56 days \citep[][]{Gaia2022b}. Assuming EIC 10 consists of two equal mass M7 stars ($0.09 M_{\odot}$), Kepler's third law places the component separation at 0.25~au, corresponding to $0\farcs023$, below the sensitivity of speckle imaging.

\citet{Salama2022} identified EIC~14 (Gaia~DR3~1097353325107339776) as a binary with an angular separation of $0\farcs4$ and a $\Delta i'$ magnitude of $2.28$, but the observation was conducted in 2012. Clark et al.\ (in prep) more recently resolved EIC~14 as a binary with an angular separation of $0\farcs29$ and $\Delta 880$\,nm magnitude of $1.62$. The angular separations at two different epochs place the binary companion between 3.5 and 5~au at a system distance of 12.3~pc.  We kept this target in our sample due to the wide separation of the binary and implemented a flux correction factor (see~\ref{subsec:ts}).

\begin{table*}[ht]
    \caption{EDEN Observations of survey targets}
    \label{tab:obsE}
    \begin{adjustwidth}{-2.25cm}{0cm}
    \resizebox{2.4\columnwidth}{!}{%
    \begin{tabular}{cccccccccc}
        \hline
        \textbf{EIC} & \textbf{Gaia DR3} & \textbf{RA (J2000)} & \textbf{Dec (J2000)} & \textbf{PM RA} & \textbf{PM Dec} & \textbf{Distance} & \textbf{Spectral} & \textbf{SpT} & \textbf{Hours}\\
        \textbf{ID} & \textbf{ID} & h:m:s & d:m:s & mas/yr & mas/yr & pc & \textbf{Type} & \textbf{Ref.} & \textbf{Obs.}\\
        \hline
        1 & 2091177593123254016 & 18:35:37.88 & 32:59:53.3 & $-72.65 \pm 0.05$ & $-755.15 \pm 0.05$ & $5.689 \pm 0.002$ & M8.5V & [1] & 204.0\\
        2 & 56252256123908096 & 3:20:59.71 & 18:54:22.8 & $352.29 \pm 0.09$ & $-257.07 \pm 0.08$ & $14.648 \pm 0.017$ & M8V & [2] & 324.8\\
        4 & 2531195858721613056 & 1:09:51.20 & -3:43:26.4 & $372.17\pm 0.27$  & $8.65 \pm 0.16$ & $10.568 \pm 0.023$ & M9Ve & [3] & 176.2\\
        5 & 31235033696866688 & 3:14:03.44 & 16:03:05.5 & $-242.96 \pm 0.19$ & $-54.85 \pm 0.16$ & $13.768 \pm 0.036$ & L0 & [4] & 100.0\\
        6 & 445100418805396608 & 3:30:48.89 & 54:13:55.2 & $-151.99 \pm 0.04$ & $3.42 \pm 0.04$ & $10.502 \pm 0.004$ & M7V & [5] & 47.9\\
        7 & 3257243312560240000 & 3:51:00.03 & -0:52:44.9 & $10.99 \pm 0.06$ & $-469.97 \pm 0.05$ & $14.678 \pm 0.015$ & M8.0V & [6] & 47.5\\
        8 & 230171768457140736 & 3:57:19.98 & 41:07:42.6 & $-204.37 \pm 0.06$ & $-24.74 \pm 0.05$ & $14.183 \pm 0.012$ & M7V & [5] & 106.4\\
        9 & 229155579195699456 & 4:19:52.13 & 42:33:30.4 & $528.67 \pm 0.08$ & $-1441.61 \pm 0.06$ & $10.262 \pm 0.010$ & M8.5V & [1] & 411.3\\
        11 & 3200303384927512960 & 4:40:23.27 & -5:30:08.1 & $334.53 \pm 0.06$ & $127.91 \pm 0.06$ & $9.747 \pm 0.006$ & M7 & [8] & 49.1\\
        12 & 3421840993510952192 & 5:10:20.09 & 27:14:01.9 & $-213.27 \pm 0.08$ & $-630.81 \pm 0.06$ & $10.276 \pm 0.007$ & M7V & [5] & 213.0\\
        13 & 191109281417914880 & 5:39:24.80 & 40:38:42.8 & $646.15 \pm 0.09$ & $-834.49 \pm 0.05$ & $11.367 \pm 0.010$ & M8.0 & [1] & 49.4\\
        14 & 1097353325107339776 & 8:25:52.82 & 69:02:01.1 & $-691.17 \pm 0.29$ & $-1276.32 \pm 0.39$ & $12.289 \pm 0.059$ & M7V & [5] & 57.0\\
        15 & 5761985432616501376 & 8:53:36.16 & -3:29:32.2 & $-516.61 \pm 0.08$ & $-199.65 \pm 0.05$ & $8.659 \pm 0.005$ & M9.0 & [1] & 65.1\\
        16 & 1227705135863076864 & 14:28:04.16 & 13:56:13.3 & $-365.42 \pm 0.06$ & $-495.51 \pm 0.06$ & $13.230 \pm 0.010$ & M7V & [5] & 57.3\\
        17 & 1287312100751643776 & 14:28:43.23 & 33:10:39.3 & $-346.96 \pm 0.04$ & $-710.99 \pm 0.09$ & $10.969 \pm 0.012$ & L0 & [9] & 219.8\\
        18 & 1282632682337912832 & 14:44:17.18 & 30:02:14.2 & $-94.39 \pm 0.07$ & $-340.33 \pm 0.07$ & $14.781 \pm 0.015$ & M8e & [10] & 80.2\\
        19 & 1262763648230973440 & 15:01:08.19 & 22:50:02.1 & $-43.12 \pm 0.11$ & $-65.14 \pm 0.14$ & $10.734 \pm 0.016$ & L0 & [9] & 64.5\\
        20 & 1272178319624018816 & 15:24:24.76 & 29:25:31.5 & $-56.77 \pm 0.03$ & $-629.24 \pm 0.04$ & $13.078 \pm 0.008$ & M7.5V & [2] & 95.7\\
        21 & 6265453524968112640 & 15:34:56.93 & -14:18:49.3 & $-918.81 \pm 0.14$ & $-330.23 \pm 0.10$ & $10.938 \pm 0.013 $ & M8.6V & [7] & 45.4\\
        22 & 4588438567346043776 & 18:26:11.00 & 30:14:18.9 & $-2290.75 \pm 0.09$ & $-683.27 \pm 0.09$ & $11.101 \pm 0.010$ & M8.5V & [11] & 80.2\\
        23 & 1762523981210977664 & 20:44:37.48 & 15:17:34.8 & $303.68 \pm 0.06$ & $-155.49 \pm 0.05$ & $10.395 \pm 0.006$ & M8 & [12] & 27.4\\
        \hline
    \end{tabular}}
    \end{adjustwidth}
    \textbf{Notes}:  TRAPPIST-1 observed in follow-up program.  Distance and proper motion values from Gaia DR3 \citep{Gaia2022a}. Spectral type references are [1]: \citet{Reiners2018}, [2]: \citet{Kirkpatrick2011}, [3]: \citet{Henry2018}, [4]: \citet{Smart2017}, [5]: \citet{Newton2014}, [6]: \citet{Deshpande2012}, [7]: \citet{BardalezGagliuffi2014}, [8]: \citet{Faherty2009}, [9]: \citet{Kiman2019}, [10]: \citet{Gizis2000}, [11]: \citet{Lepine2002}, [12] Photometric type from \citet{Reyle2018}.
\end{table*}

\begin{table*}[ht]
    \begin{adjustwidth}{0.65cm}{0cm}
    \caption{Apparent magnitudes of the EDEN targets presented in this study.}
    \label{tab:mags}
    \begin{tabular}{ccccccccc}
    \hline
    \textbf{EIC} & \textbf{\textit{G}} & \textbf{e}$_{\mathbf{\mathit{G}}}$ & \textbf{\textit{J}} & \textbf{e}$_{\mathbf{\mathit{J}}}$ & \textbf{\textit{H}} & \textbf{e}$_{\mathbf{\mathit{H}}}$ & \textbf{\textit{K}} & \textbf{e}$_{\mathbf{\mathit{K}}}$ \\
    \textbf{ID} & $\mathrm{mag}$ & $\mathrm{mag}$ & $\mathrm{mag}$ & $\mathrm{mag}$ & $\mathrm{mag}$ & $\mathrm{mag}$ & $\mathrm{mag}$ & $\mathrm{mag}$ \\
    \hline
    1 & 14.8510 & 0.0029 & 10.270 & 0.022 & 9.617 & 0.021 & 9.171 & 0.018 \\
    2 & 16.1048 & 0.0032 & 11.759 & 0.021 & 11.066 & 0.022 & 10.639 & 0.018 \\
    4 & 16.3750 & 0.0031 & 11.694 & 0.021 & 10.931 & 0.026 & 10.428 & 0.025 \\
    5 & 17.2427 & 0.0035 & 12.526 & 0.024 & 11.823 & 0.036 & 11.238 & 0.021 \\
    6 & 13.7143 & 0.0028 & 10.173 & 0.021 & 9.595 & 0.016 & 9.284 & 0.018 \\
    7 & 15.3131 & 0.0028 & 11.302 & 0.024 & 10.609 & 0.022 & 10.232 & 0.024 \\
    8 & 14.6381 & 0.0029 & 10.903 & 0.023 & 10.299 & 0.020 & 9.949 & 0.019 \\
    9 & 15.5551 & 0.0028 & 11.094 & 0.022 & 10.383 & 0.028 & 9.900 & 0.022 \\
    11 & 14.9043 & 0.0029 & 10.658 & 0.024 & 9.986 & 0.022 & 9.545 & 0.019 \\
    12 & 14.9257 & 0.0030 & 10.698 & 0.020 & 9.965 & 0.022 & 9.560 & 0.018 \\
    13 & 15.2733 & 0.0029 & 11.109 & 0.021 & 10.446 & 0.021 & 10.044 & 0.018 \\
    14 & 13.7621 & 0.0038 & 10.078 & 0.021 & 9.496 & 0.019 & 9.161 & 0.014 \\
    15 & 15.8889 & 0.0030 & 11.212 & 0.026 & 10.469 & 0.026 & 9.942 & 0.024 \\
    16 & 14.8963 & 0.0029 & 11.014 & 0.021 & 10.390 & 0.015 & 10.026 & 0.020 \\
    17 & 16.6346 & 0.0028 & 11.990 & 0.021 & 11.225 & 0.029 & 10.744 & 0.024 \\
    18 & 15.9061 & 0.0029 & 11.671 & 0.020 & 11.017 & 0.021 & 10.616 & 0.018 \\
    19 & 16.5353 & 0.0030 & 11.866 & 0.022 & 11.181 & 0.030 & 10.706 & 0.024 \\
    20 & 15.2801 & 0.0028 & 11.206 & 0.022 & 10.535 & 0.021 & 10.155 & 0.015 \\
    21 & 15.7637 & 0.0029 & 11.380 & 0.023 & 10.732 & 0.022 & 10.305 & 0.023 \\
    22 & 15.9470 & 0.0030 & 11.659 & 0.020 & 11.175 & 0.016 & 10.811 & 0.019 \\
    23 & 15.2678 & 0.0028 & 11.025 & 0.022 & 10.490 & 0.016 & 10.061 & 0.018 \\
    \hline
    \end{tabular}
    \end{adjustwidth}
    \textbf{Notes}: Gaia magnitudes from DR3 \citep[][]{Gaia2022a}, JHK magnitudes from 2MASS \citep[][]{Skrutskie2006}
\end{table*}

The sample of stars matching our criteria (including the wide-separation binary) is shown in Tables~\ref{tab:obsE} and~\ref{tab:mags}.  As the completeness of our targets within 15~pc neared 100\% during the survey, we extended our observations to stars in the 15--21~pc range, as well as many targets outside the criteria (either in distance or in spectral type) as interesting candidates during the time when no $<$15~pc target was seasonally observable, as well as interesting candidates for other scientific purposes \citep[e.g., the high-frequency flare star Wolf 359;][]{Lin2021}.  These targets are shown in Table~\ref{tab:obsNE}.  In addition, we performed targeted follow-up observations of exoplanet candidate stars found by TESS (see Table~\ref{tab:obsFOP}).

\section{Data Reduction} \label{sec:reduc}

Here we provide a summary of our photometric pipeline \texttt{edenAP} and additional post-processing steps we took to prepare the data for our transit search.

\subsection{Photometric Pipeline} \label{subsec:AP}

We began our photometric process by calibrating the target frames using combined flats and biases. This step is coded to be optional in case of bad or unavailable calibration files, which only rarely occurs ($<5\%$ of nights) and does not significantly reduce the quality of our light curves.  We then performed aperture photometry using \texttt{photutils} \citep[][]{Bradley2019}, including calculating an astrometric solution using \texttt{astrometry.net} \citep[][]{Lang2010}. We selected the aperture size between 5-50 pixels in steps of 1 pixel that minimizes the RMS scatter in the light curve of the target and a sky background median value of a 60x60 pixel box around the target after other sources are removed.  Finally, we ran a simple comparison detrending of the light curve.  The six reference stars in the field (or as many as possible in the rare cases where there are few reference stars) with the lowest average deviation within bins of 20 exposures across the light curve were median-combined into a single unbinned light curve, which was then divided out from the target light curve \citep[see e.g.,][Section 3.4]{Gibbs2020}.  Further detrending was performed as part of the transit search, as described in the following section.

\subsection{Light Curve Detrending} \label{subsec:detrend}

As is usual for ground-based, high-precision photometry, our data are impacted by relatively slowly changing systematics, usually arising from airmass changes, telescope positional drift, and seeing variations.  In order to correct for these systematics, we performed detailed detrending analysis similar to that laid out in Section 4.2 of \citet[][]{Gibbs2020}.  Specifically, we followed these steps:
\begin{itemize}
    \item We divided the normalized light curve by a 2nd-order polynomial that we fit to the original data.
    \item We performed a median filter over a 2 hour window.
    \item We removed any data points 2$\sigma$ above the median but not below to avoid removing transits much shorter (1-2 hours) than the length of the overall light curve ($>$3 hours).
    \item We fit a 2nd-order multivariate polynomial, based on the airmass, background level, and pixel x-y positions and divided it out.
\end{itemize}

Although we adopted steps from \citet[][]{Gibbs2020}, we also explored whether alternative detrending approaches may yield better results. Specifically, we tested multiple detrending procedures from the \texttt{W\={o}tan} package \citep[][]{Hippke2019b}, including combinations of biweight filtering, median filtering, spline fitting, and Savitsky-Golay polynomial fitting.  For injected planet transit tests, none of the above versions of detrending procedures provided a lower scatter in the out-of-injected-transit baselines without also overfitting and at least partially removing the injected transit from the dataset.  Specifically, the biweight filtering is best at removing signals from the raw dataset that have much longer timescales than the transit signal. For short-timescale signals across the duration of one observing night, biweight filtering did not improve the results.  Therefore, our tests did not demonstrate that alternative approaches result in significant improvement, and we thus decided to continue using our original detrending procedure.

\subsection{Variable Precipitable Water Vapor}\label{sec:PWV}

A potential source for false positive detections are variations of precipitable water vapor (PWV) in Earth's atmosphere, in particular for red target stars such as the ones in our survey~\citep[e.g.,][]{Bailer-Jones2003,Blake2008}. These variations may affect the accuracy of photometric time series in magnitudes and time scales that could be comparable to transit signals~\citep{Berta2012}.

The filter bands in which EDEN observations are done are sometimes affected by PWV, as they overlap with several water vapor bands. The magnitude of the light curve impairment due to PWV depends on the spectral energy distributions of the target and comparison stars, the amount of water vapor in the atmospheric column along the line of sight of the telescope and its variation in time, and the bandwidth and wavelength of the used filters. Mitigation strategies have been put forward, such as evaluating the contribution of PWV across global, all-sky light curves and/or monitoring using local environmental sensors \citep[][]{Pedersen2023}, as well as concurrent satellite remote sensing observations to drive a correction \citep[][]{MeierValdes2021}.

Given the above, PWV — if it affects LCs significantly – is expected to introduce false positives. However, there is no evidence for such effect in our dataset: the fact that we only detected one transit candidate (see Sect.~\ref{subsec:J0419+4233}) in $> 2450$~hours of photometry shows that PWV is not a major source of false positives for the combination of filter choice and telescope locations of the EDEN survey. However, in the case of a signal that survives initial validation steps, we will need to evaluate the possibility of a false positive scenario caused by PWV.

\section{Analysis} \label{sec:analysis}

\subsection{Transit search} \label{subsec:ts}

We first analyzed all the data we had on each target by eye via the EDEN Interactive Viewer, which shows the light curve for the night along with airmass, moon altitude, sky background, reference and target star flux, and x-y pixel positions.  Transit-like features were manually flagged for potential follow-up, but most variations in the data can be attributed to observing conditions or stellar activity. We account for the dilution of potential transit depths from the binary star EIC 14 by implementing a correction factor following Equation~4 from \citet[][]{Furlan2017}, with $\Delta m = 1.62$). We note that our GG495 long-pass filter is not the same as the narrow-band 880\,nm of the speckle observations, making this $\Delta m$ value an approximation. Assuming a planet orbits the primary, brighter star in the system, the correction factor is $\sim$1.1.

To more thoroughly search for signatures of planetary transits in EDEN photometry, we used the Transit Least Squares \citep[\texttt{TLS};][]{Hippke2019a} search algorithm, which is based on a template signal shape informed by small-planet detections of the \textit{Kepler} mission. We applied the algorithm on all stars in our sample with data cuts at 0.3\%, 0.6\%, 0.9\%, and 10\% precision levels. Each of these runs returns a periodogram, model light curve (including phase folding), and best-fit parameters (period, duration, depth/radius ratio, mid-transit time for first transit in our data, and false alarm probability) for the expected transits.  We note that \texttt{TLS} always attempts to fit a transit model, and that many of the output parameters are likely not actual transits but instead false positives that are the result of attempting to fit a transit to short-period and short-duration noise residuals in the light curves.  We vetted all the signals and discarded any transit models with high false alarm probability ($>0.01$) as well as those signals with low false alarm probabilities but with parameters for planets that are unlikely to exist and be detected with our observations: transit durations $\lesssim$15 minutes, orbital periods $<0.25$ days, transit depths corresponding to planets $< 0.1 R_\oplus$.

\subsection{Detection sensitivity} \label{subsec:sense}

In order to probe the underlying planet population through our detection statistics, a thorough understanding of our survey completeness is necessary.  Besides the geometric probability of a planet to transit given our line of sight to its host star, the completeness of our survey is limited by our ability to detect transit signatures.  Our sensitivity is mostly limited by the photometric precision of our light curves and by the performance of our detection pipeline.

For close-to-ideal surveys not suffering from non-uniform sampling and significant data gaps, analytic expressions exist to estimate the sensitivity for transit signals \citep[e.g.,][]{Pepper2003}. 
In the case of our ground-based survey with its inevitable irregularities, we instead employed injection-and-retrieval simulations. Our approach was to inject planet signatures on a grid in the orbital period and planet radius and recover them.  We used a log-spaced grid of 20 orbital period bins from 0.5-10 days, but if we had less than 10 nights of data on a target we truncated the grid to the number of nights we had.  We also used a log-spaced grid of 12 planet radius bins from $0.6-3.5\,R_\oplus$.  For the targets with data covering less than 10 nights, we artificially cut off the injections to periods where we would see at least two transits of the injected planet.

\begin{figure*}
    \centering
    \includegraphics[width=\linewidth]{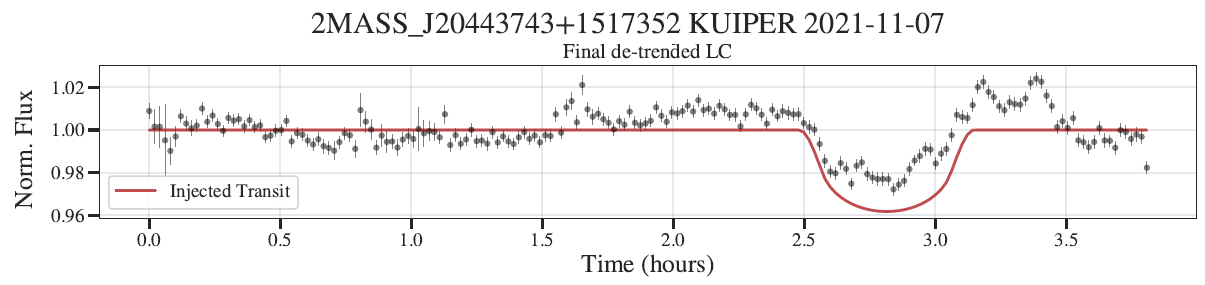}
    \caption{A detrended light curve with an injected transit from the Kuiper 61'' telescope on 07 Nov 2021 UT.  The transit is deep enough that the detrending raises the baseline pre- and post-transit slightly compared to the rest of the light curve, but this does not affect our ability to recover the injected planet because it is such a strong signal (SNR $\sim$ 6).}
    \label{fig:lcit}
\end{figure*}

We created 250 planets with a random orbital period, orbital phase, and planet radius within each of the given grid bins, where the random values were drawn from a uniform distribution within each bin. This differs from the process in \citealt[][]{Gibbs2020}, where they injected enough planets to retrieve 10 observable transiting planets per bin. This is due to increased noise levels generated by the previous method (see Figure~5 in \citealt[][]{Gibbs2020}). In addition, we assumed the orbits to be circular, we randomly drew the impact parameter from a uniform distribution between 0 and 0.7 to avoid grazing transits, and we used the \citet[][]{Claret1998} limb-darkening law with coefficients $u_1=0.84$ and $u_2=0.125$.  
We injected the transit signals of each artificial planet into our observed photometric time series, detrended the resulting light curve, and ran the \texttt{TLS} algorithm to attempt to recover the injected planets.  An example detrended light curve with an injected transit is shown in Figure~\ref{fig:lcit}. The results of our transit search are described in Section~\ref{subsec:res_ts}.

We considered our injected planets to be detected and positively recovered if the period was within 0.02 days (or 24 minutes) of the injected period, which corresponds to a fractional difference $\leq 4\%$ based on the actual given period. We also accounted for undersampled periods due to the scarcity of our data by looking at light curves with at least 2 transits found and checking to see if the given period for the injected planet still matched the injected transits recovered. We created a separate detection filter for single transits found and discarded those, as we cannot place our desired constraint on the orbital period of a planet with only one transit recovered.

\section{Results} \label{sec:results}

\subsection{High-precision light curves for the EDEN Targets}

Our survey yielded a rich dataset of detrended light curves with $\sim1-10 \, \mathrm{mmag}$ scatter for \ntargets{} EDEN target stars ranging in spectral types from M7V to L0, and in distance from 5.7~pc to 14.8~pc.  For most target stars, our dataset provides the highest quality photometric monitoring to date, often with a unique combination of relatively high cadence ($\sim$60 seconds), sensitivity, and monitoring baseline. The light curves were used in this study to search for transiting exoplanets and place upper limits on their occurrence rates, but they can also be exploited for studies of stellar activity and stellar rotation, among other uses \citep[see e.g.,][]{Lin2021, Murray2022}. Although not included here, our dataset also provides similar photometric monitoring information for typically a dozen other stars in the same field of view of our target stars.






\subsection{Transit Search} \label{subsec:res_ts}

Our careful manual and algorithmic analyses of the light curves resulted in no convincing transit detections.  Roughly 20 potential transit signals were flagged in all of our data, and there was one repeating, transit-like signal, which we followed up with additional observations and found no further evidence of a transit (see Section~\ref{subsec:J0419+4233}).  Many \texttt{TLS} results were discarded as being unphysical false positives via the vetting procedure mentioned in Section~\ref{subsec:ts}. Therefore, we conclude that \texttt{TLS} found no real transit signals in the data from any of our targets.  The fact that our light curves did not show many false positives --- yet remained very sensitive to exoplanets down to Earth-radii --- demonstrates the efficiency of our data reduction and analysis approach.
\subsection{Detection sensitivity}

\begin{figure*}
    \centering
    \includegraphics[width=0.845\linewidth]{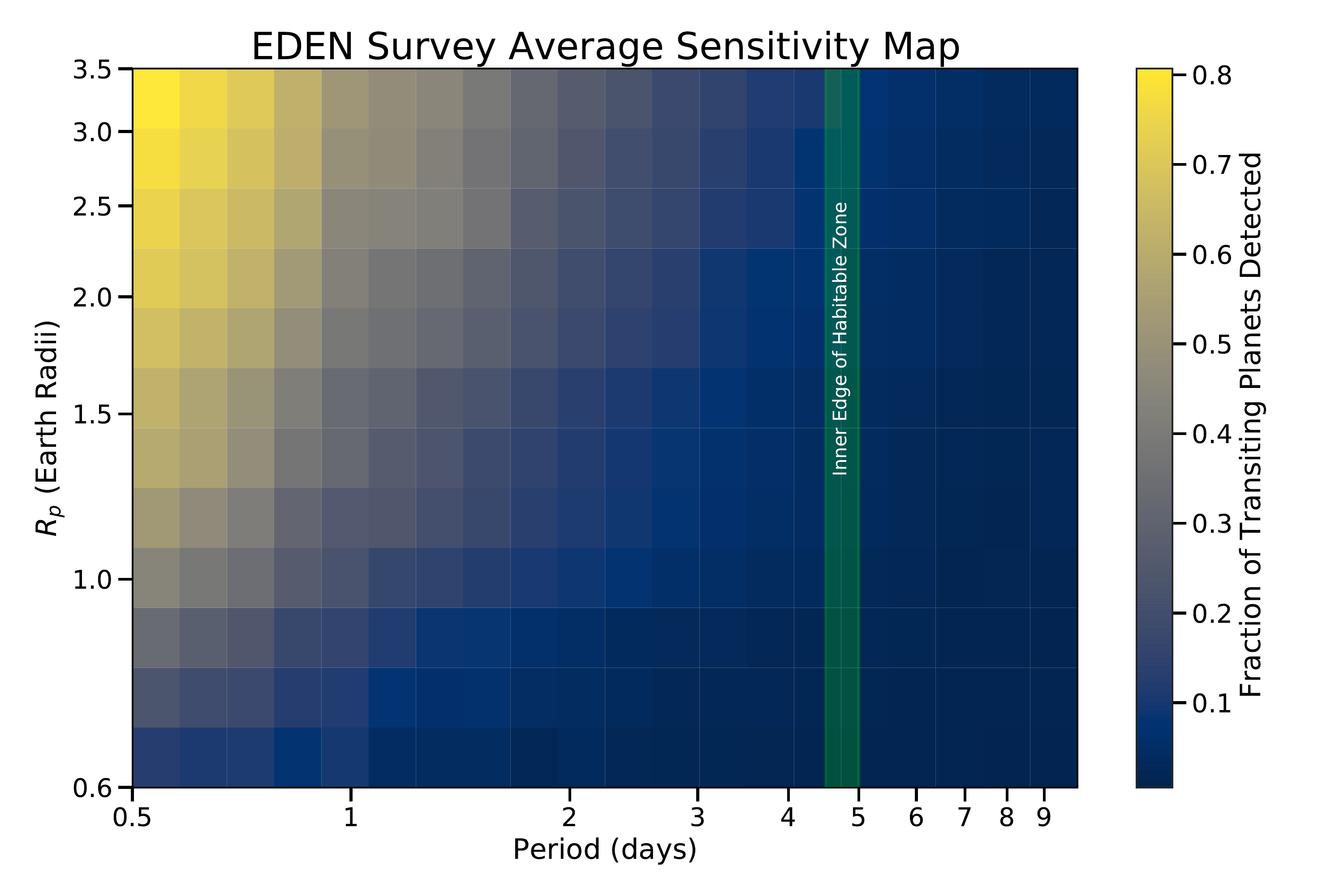}
    \includegraphics[width=0.845\linewidth]{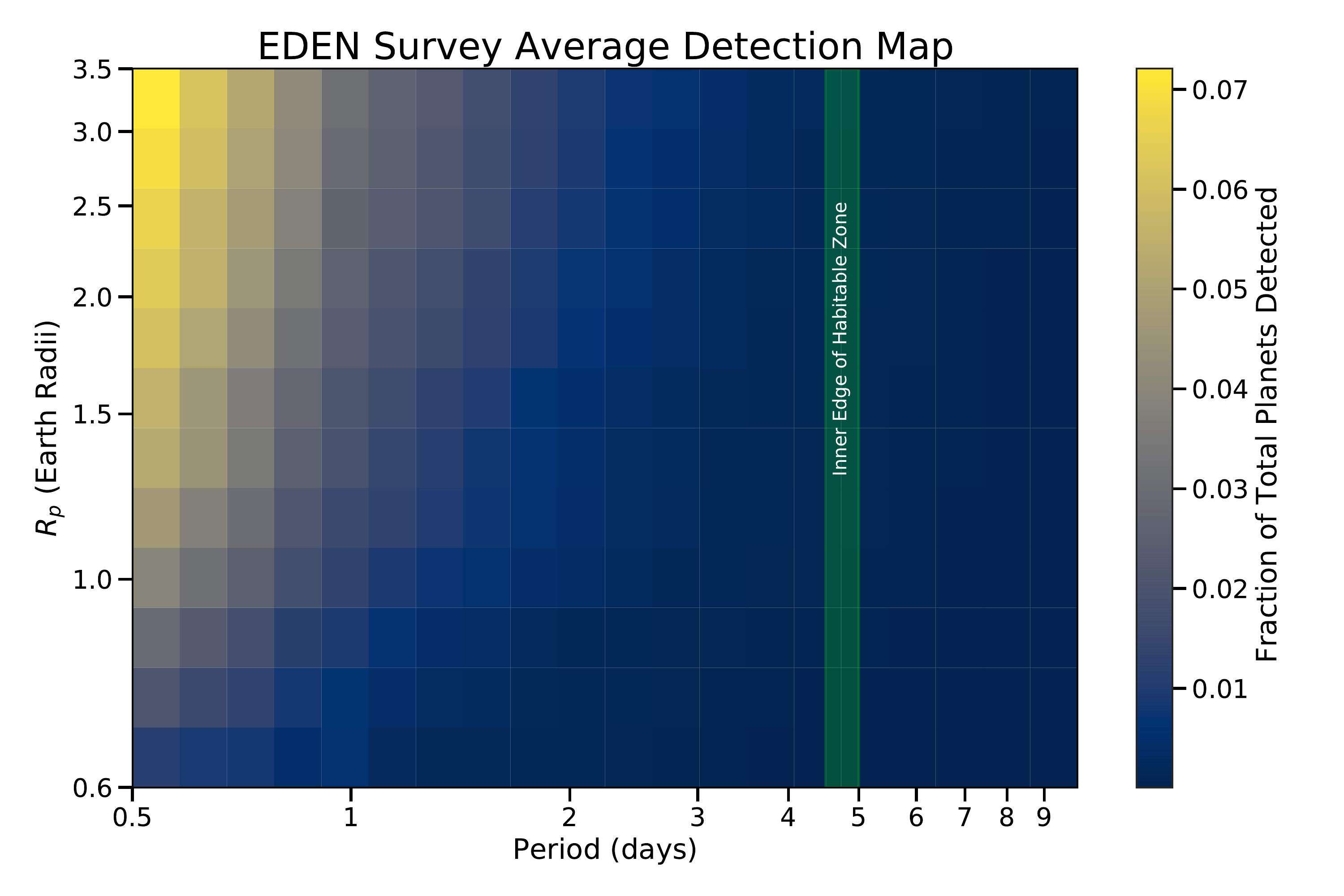}
    \caption{The average sensitivity (top) and full detection probability including geometric transit likelihood (bottom) of the EDEN survey for planets with periods between 0.5 and 10 days and sizes between 0.6 and 3.5 Earth radii.  For ultra-short-period planets ($<$ 1 day), we reach 50\% sensitivity for planets with radius $1.5 R_\oplus$ or larger.  Due to the low geometric transit probability, the overall completeness is much lower and drops sharply with the orbital period.  The inner edge of the habitable zone is defined using the \citet[][]{Kopparapu2013} model based on the temperature and insolation (assuming a stellar effective temperature of 2600 K and a planetary albedo of 0.3).}
    \label{fig:map1}
\end{figure*}

We report the percentage of planets recovered with properties similar to the injected ones and created a map of the recovery probability in our sensitivity maps (see Figure~\ref{fig:map1} for the average map and full detection probability from the EDEN survey and Appendix~\ref{sec:maps} for the full set of 22 sensitivity maps).

Our average sensitivity to approximately Neptune-sized planets ($R\sim3.5 R_\oplus$) on very short ($P\sim0.5$~days) orbits for most of our target stars is $\sim85-90\%$ (see the top left corner of the top panel of Figure~\ref{fig:map1}), which can be extended as a conservative estimate to larger planets as well.  Due to the nightly cadence of observations, we tend to see a higher sensitivity for our targets at periods that are integer multiples of one day.  Overall, our light curves are very sensitive to short-period ($P <1.5$~days) super-Earths and larger ($R >1.5R_\oplus$) exoplanets and, for most of our targets, we can exclude that such transiting planets orbit our targets.   

We note that EIC 7 (2MASS J05310004-0052452) and EIC 21 (2MASS J15345704-1418486) have significantly lower sensitivity than the other targets, which occurs due to a combination of low overall time observed (<50 hours), a southern declination, and relatively high light curve scatter for more observations than other targets. However, these two targets do not significantly affect the overall average sensitvity map, and we include them here for completeness.

\section{Population-Level Insights}\label{sec:hypothesis_testing}

Our well-characterized sensitivity and relatively large sample size allow us not only to provide robust upper limits in a previously poorly explored region of the parameter space, but also to contrast the outcome of our survey (no planet detection) to different hypothetical planet populations. In the following analysis, we explored two demographic features: First, we analyzed the occurrence rate of TRAPPIST-1~b analogs and TRAPPIST-1-like planetary systems. Second, we analyzed the occurrence rate of close-in ($P\lesssim1$ day) giant planets.

In both cases, we used the survey completeness described in Section~\ref{subsec:sense} to compare our results to a hypothetical planet population. This approach follows the Monte Carlo assessment method introduced in \citet[][]{Kasper2007} to contrast non-detections from a direct imaging survey with possible extrapolations of the exoplanet population detected by radial velocity surveys to larger separations. Since then, this method has been used for a variety of similar applications \citep[e.g.,][]{Mulders2018}.

This approach -- forward-modeling and quantitative comparison to the predictions -- is generally preferable to the inversion of sensitivity maps to occurrence rates for two reasons: (1) it allows for a more accurate consideration of sensitivity differences between individual targets and for a better handling of the subtleties of the observational biases; and (2) it provides more robust results for parameter ranges with only a few detections. For a detailed discussion of the advantages of this approach, see \citet[][]{Mulders2018}.

In this implementation of the Monte Carlo assessment, we simulated observations of synthetic planets in the measured EDEN light curves, with the planet parameters drawn from model distributions corresponding to the specific hypothesis tested - for example, a TRAPPIST-1~b-like planet would be a 1.1 $R_\oplus$ planet orbiting at P = 1.5 days with a random inclination.  We then determined in what fraction of the simulated surveys the number of detections would be consistent (at $2\sigma$) with the outcome of the EDEN survey. For example, if our survey resulted in zero planet detections, but our hypothesis would lead to planet detections in 95\% of the simulated surveys, we would be able to exclude that hypothesis at a 95\% confidence level.

\begin{figure*}
    \centering
    \includegraphics[width=\linewidth]{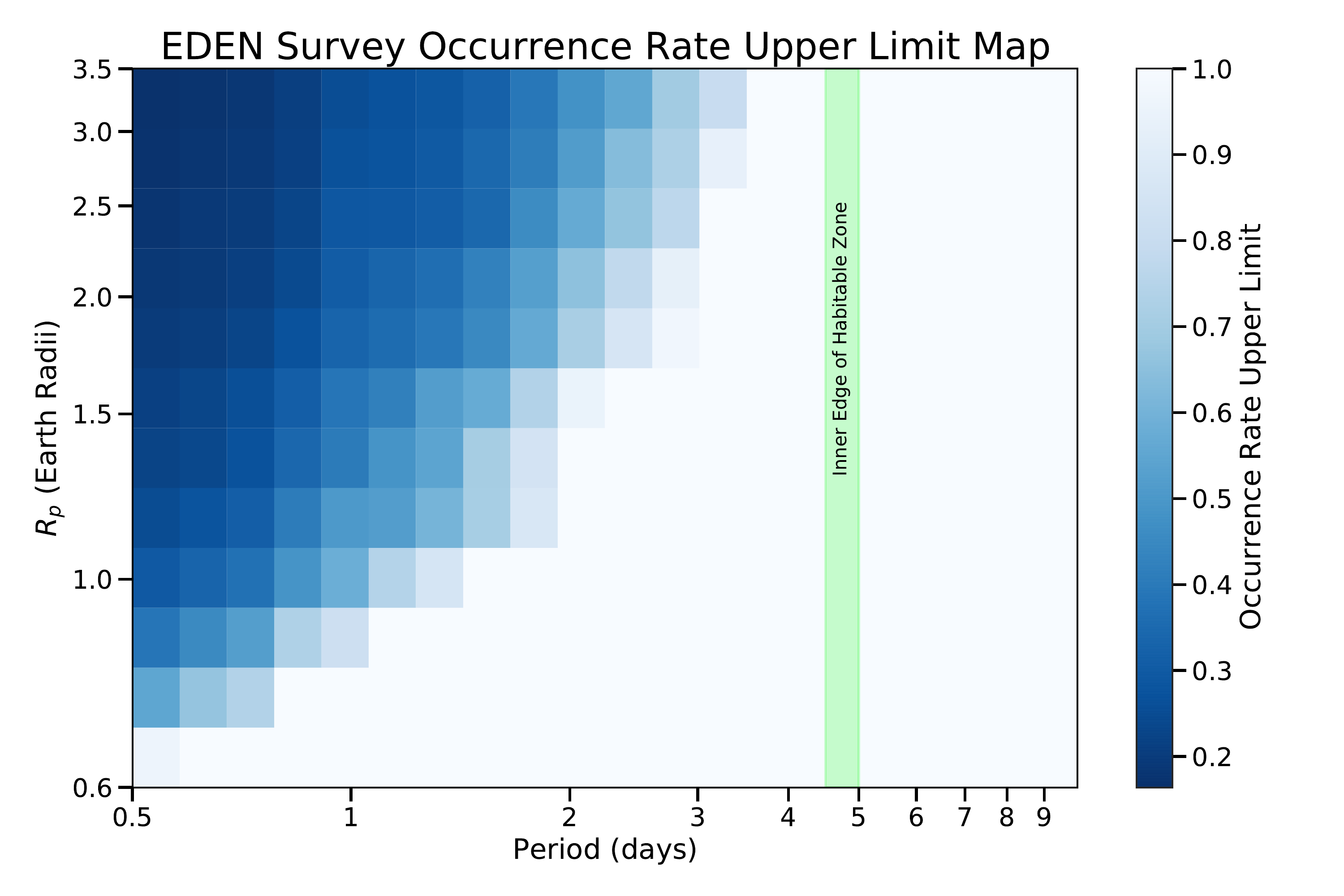}
    \caption{The upper limit on the occurrence rate from our analysis that follows the procedure from \citet[][]{Sagear2020}, assuming $P_{null} = 0.05$. We find similar occurrence rates when adjusting for bin size and type (log-linear vs log-log in orbital period and planet radius) for, e.g., mini-Neptune planets with orbital periods just greater than 1 day.}
    \label{fig:ulim}
\end{figure*}

We also performed an upper-limit analysis for the occurrence rates following the procedure by \citet[][]{Sagear2020}. We set the probability of detecting zero planets to a given value (here 0.05) and then calculated the occurrence rate necessary to reach that null detection probability, given our detection sensitivities for each target star \citep[see Equation 3 in][]{Sagear2020}. The results from our analysis are shown in Figure~\ref{fig:ulim}. We find that we have similar results to the upper limits from \citet[][]{Sagear2020} when adjusting for their different bin sizes and types (log-linear vs log-log in orbital period and planet radius) for, e.g., mini-Neptune planets with orbital periods just greater than 1 day. We also provide detailed upper limits for short-period ($< 1$ day) planets down to 0.6 $R_\oplus$. These values are also similar to the occurrence rates from \citet[][]{SestovicDemory2020}. While the K2 occurrence rates were indeed slightly more limiting, they came from a biased sample of M dwarfs chosen by users instead of a volume-complete sample, so our work complements and extends that of previous studies to this nearby sample.

\subsection{The Occurrence Rate of TRAPPIST-1~b Analogs and TRAPPIST-1-like Planetary Systems}

Using the above-described Monte Carlo method to statistically compare the results of simulated surveys of hypothetical exoplanet populations to the actual results of our survey, we first tested the hypothesis that ``TRAPPIST-1~b analogs orbit every late M~dwarf star'', or $f_\mathrm{Trb} = 100\%$. Although simplistic, this hypothesis is a sensible starting point for the exploration of the exoplanet population around late M~dwarfs, since short-period planets like TRAPPIST-1~b are among the easiest to detect in the known late M~dwarf population.

In order to do so, we simulated our EDEN transit survey 10,000 times to determine in what fraction of these surveys would the predicted results be consistent with our actual survey.  Assuming isotropic orbit orientations, we injected a planet with a probability $P_\mathrm{geom} = (R_\star+R_p)/a$, which approximates the geometric transit probability for a planet with radius $R_p$ on a circular orbit with semi-major axis $a$ around a star with radius $R_\star$. In the case of a transit occurring, we injected the photometric signature of a planet with random orbital phase and \mbox{TRAPPIST-1~b's} orbital period and planet radius \citep[P=1.511 days, R=1.116 $R_\oplus$;][]{Agol2021} and tested its detection by our pipeline. In 15\% of our simulated surveys, we would have seen at least one planet out of our sample of \ntargets{} stars.  Our sensitivity (probability to detect a transiting TRAPPIST-1~b) is $\sim18\%$, whereas the completeness (probability of detecting a planet, so the sensitivity multiplied by the geometric transit probability) is $\sim0.8\%$.

We also tested the hypothesis that ``TRAPPIST-1-like planetary systems orbit every late M~dwarf star'' ($f_\mathrm{Tr} < 100\%$) as an extension to the TRAPPIST-1~b hypothesis.  The additional planets in the full TRAPPIST-1 system make the system as a whole more likely to be discovered, even with individual planets decreasing in detection likelihood going outwards in the system.  We took all of the planet period and radius values from \citet[][]{Agol2021}, and set a random system inclination and a Rayleigh distribution for mutual inclinations around $2^\circ$ \citep[e.g.,][although \citeauthor{Luger2017} \citeyear{Luger2017} measured the TRAPPIST-1 mutual orbital inclinations to be around 0.3$^\circ$]{Fabrycky2014}. Planets g and h were included, but with the survey sensitivity as if they were at an orbital period of 10 days because their orbital periods are beyond 10 days. Again the EDEN transit survey was simulated 10,000 times, and in 21\% of the surveys we would have detected at least one planet.

\subsection{Giant Planet Occurrence Rates}\label{sec:hypotest_giants}
In the following, we provide an estimate of the occurrence rate of close-in giant planets around late M~dwarfs based on our survey's completeness in this parameter domain.  Our constraint is based on the same numerical Monte Carlo experiment as above for TRAPPIST-1 b analogs. As a conservative approximation, we assumed for these planets the sensitivity we found for the largest artificial planets that we injected in our sensitivity analysis (\SI{3.5}{R_\oplus}, see Section~\ref{subsec:sense}). For a host star radius representative of an M8V dwarf~\citep[$R_\star = 0.114 \, R_\odot$;][]{Pecaut2013}, the geometric transit probability $P_{geom} \approx \SI{11}{\percent}$ for a planet on a \SI{1.05}{\day} orbit. Even assuming such close orbits, \SI{98}{\percent} of our mock surveys yield zero detections if giant planets occur  around \SI{2}{\percent} of late M~dwarfs~\citep[][]{Ghezzi2018}. If instead every such star hosts a giant planet, we would expect to detect at least one giant planet in our survey in \SI{73}{\percent} of the cases.  If we decrease the orbital period down to \SI{0.5}{\day} orbital periods, the average sensitivity to a transiting planet increases by a factor of $\sim1.6$, and we can constrain the giant planet occurrence rate down to \SI{75}{\percent} with 95\% confidence.

\section{Discussion} \label{sec:discuss}

\subsection{Light curves}

Our EDEN observations provide the most extensive set of light curves for the majority of our \ntargets{} target stars. Table~\ref{tab:obsE} summarizes our targets and the total length of observations for each. Even though our targets are close to the Solar System ($d<$\EDENVolume{}), the majority of the target stars are too faint for wide-array or all-sky surveys -- such as HAT \citep[][]{Bakos2002}, MEarth \citep[][]{Irwin2009}, or TESS \citep[][]{Ricker2015} -- to search for planets. For example, a comparison of our targets (with typical spectral types of M7--M9) to the stellar effective temperature and planet radius distribution of TESS Objects of Interest (TOI) as of April 2022 (see Figure.~\ref{fig:TESSTOIST}) illustrates how well our observations complement the TESS survey: there are currently no TOIs with spectral types beyond M7. This paucity of TOIs for late spectral types is primarily due to the fact that TESS's sensitivity -- in spite of extreme photometric stability -- is limited to brighter stars by its small collecting area. The MEarth project, which utilized arrays of small robotic telescopes to search for transiting planets around bright mid-to-late M~dwarfs~\citep{Nutzman2008}, reports photometric precisions of \SIrange{0.2}{1.0}{\percent} at a typical cadence of $\sim \SI{25}{\minute}$~\citep{Berta2013}.

In comparison to these other current M~dwarf surveys, the EDEN survey, due to the large apertures and uniform observing and data reduction strategy, accomplishes both uniquely long monitoring with high cadence and high sensitivity. Specifically, our targets are observed for at least \avgtime{} (the equivalent of about four typical, clear nights) and 1/3 of our targets were observed for over 100~hours (the equivalent of about 15 typical, clear nights) with $\sim$60 second cadence.  As the light curves collected in our EDEN survey will likely be of use for a variety of future studies (e.g.,  stellar activity and rotation measurements), we make the reduced and calibrated light curves available to the community as an online data set accompanying this paper. The non-EDEN targets observed in our survey (Tables~\ref{tab:obsNE} and \ref{tab:obsFOP}) represent a less homogeneous dataset. Those light curves and the ensemble of images obtained are available on a collaborative basis and will be made fully available in the near future.

\subsection{A possible signal on EIC 9 (2MASS~J04195212+4233304)} \label{subsec:J0419+4233}

\begin{figure*}
    \centering
    \includegraphics[width=0.68\linewidth]{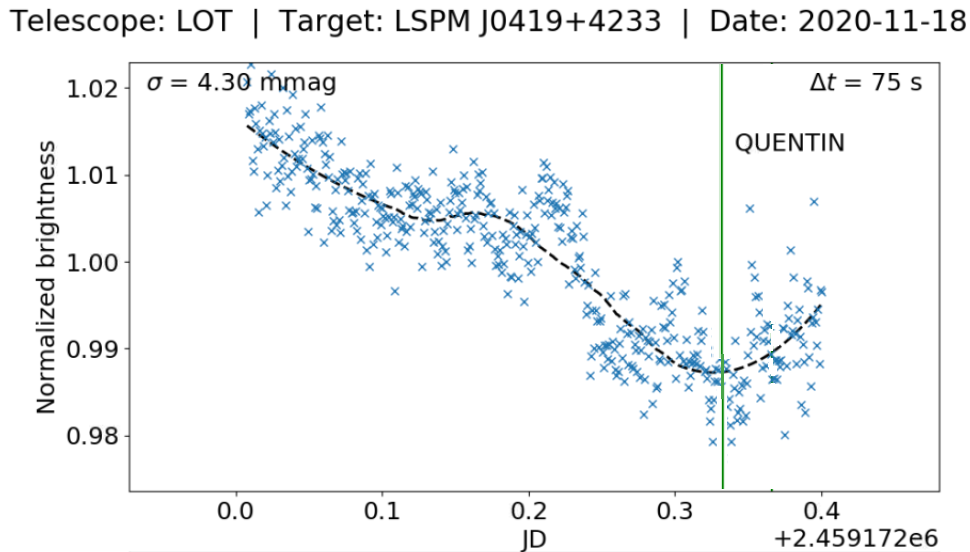}\\
    \includegraphics[width=0.65\linewidth]{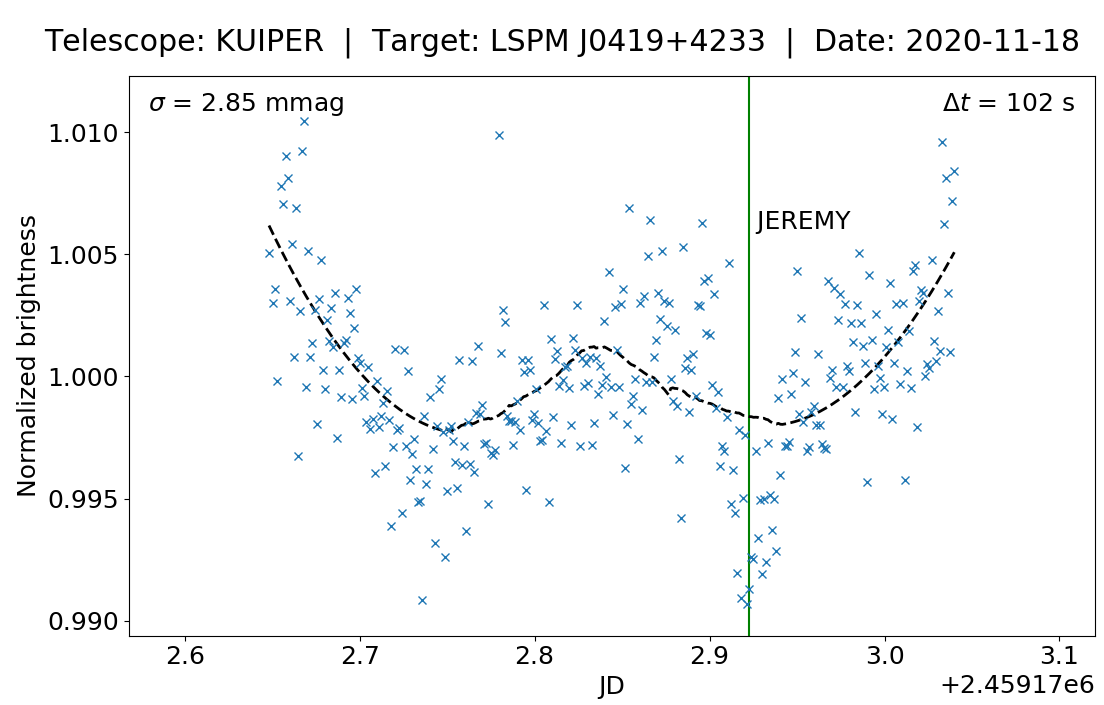}\\
    \includegraphics[width=0.65\linewidth]{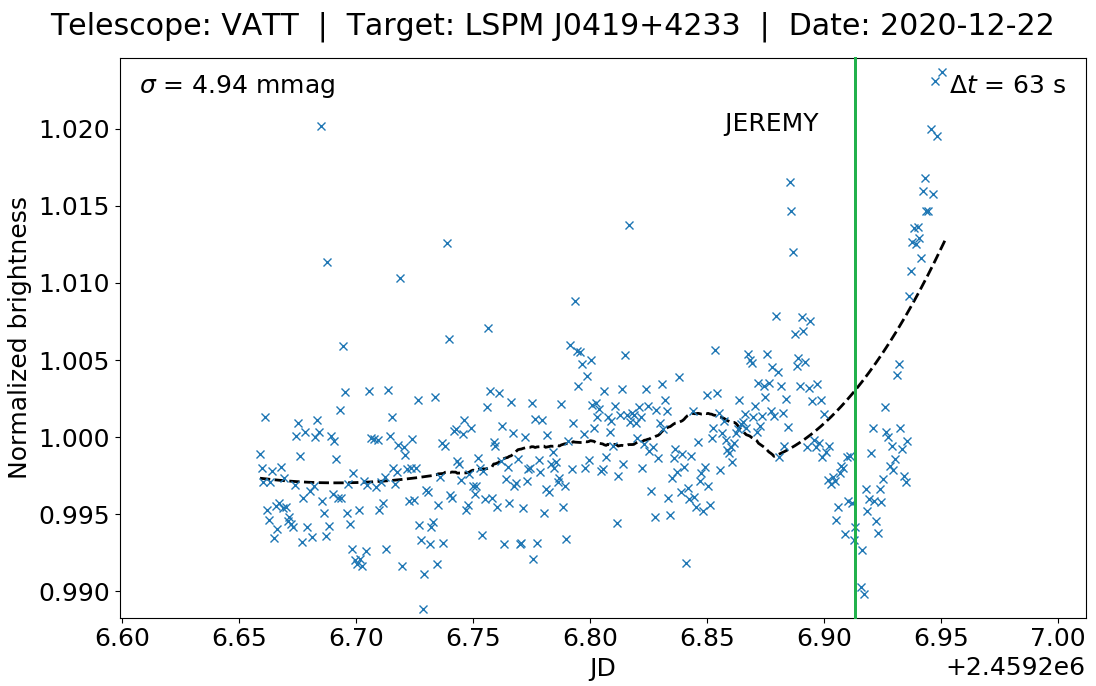}
    \caption{The first look at the non-detrended light curves from a simple reference star calibration of each potential transit event, before full detrending is done. The transit-like features are marked with green lines and the name of the Project EDEN member who analyzed the quick look data.}
    \label{fig:EIC9FL}
\end{figure*}

\begin{figure*}
    \centering
    \hspace{0.85cm}\includegraphics[width=0.75\linewidth]{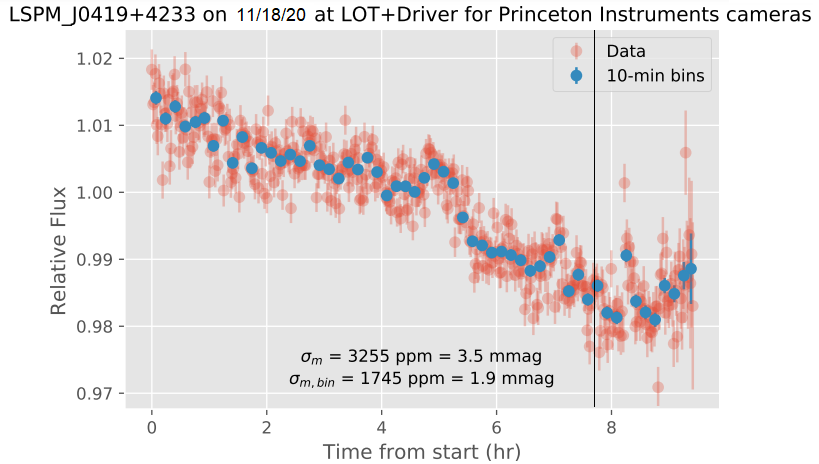}\\
    \hspace{-0.3cm}\includegraphics[width=0.60\linewidth]{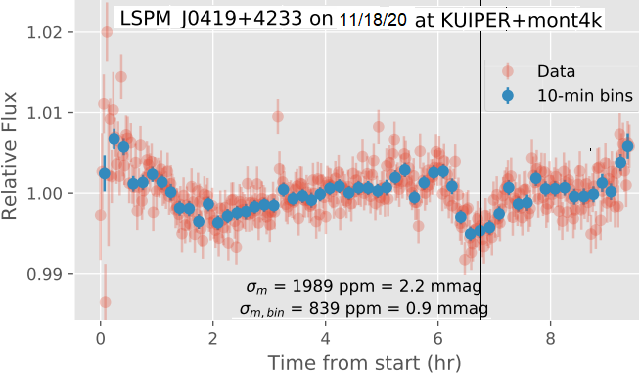}\\
    \hspace{-1cm}\includegraphics[width=0.54\linewidth]{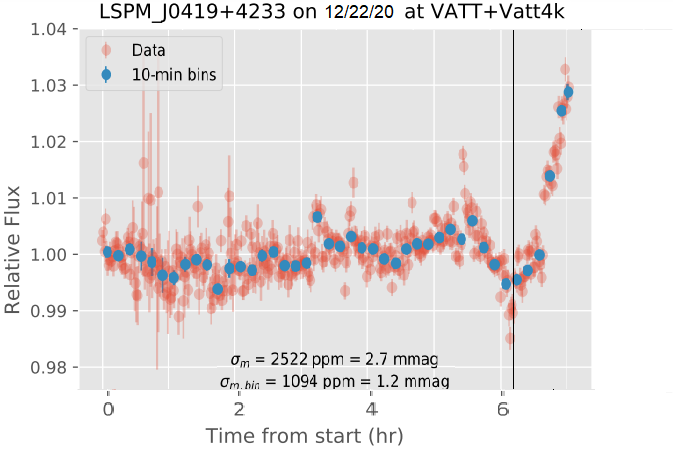}
    \caption{The three light curves (both full and 10-minute binned data) with errorbars for EIC 9 where we identified transit-like features (noted by black vertical lines in each panel). Top: Light curve from LOT on 18 Nov 2020 (UT); transit event originally identified (see Figure~\ref{fig:EIC9FL}) centered at 7.8 hours (2459172.33 JD). Middle: Light curve from Kuiper on 18 Nov 2020 (UT) with transit event centered at 6.8 hours (2459172.91 JD). Bottom: Light curve from VATT on 22 Dec 2020 (UT); transit event centered at 6.1 hours (2459206.91 JD.)}
    \label{fig:EIC9LC}
\end{figure*}

During observation runs in fall-winter 2020, we identified transit-like features in light curves of EIC 9 (2MASS~J04195212+4233304), a 0.5-1\% dip with a duration of $\sim$45 minutes.  Similar features occurred twice on the same day of 18 Nov 2020 (UT), once observed by the Lulin One-meter Telescope (LOT) in Taiwan and again about 14 hours later by the Kuiper 61" Telescope in Arizona. The Vatican Advanced Technology Telescope (VATT) in Arizona observed another feature on 22 Dec 2020 (UT).  The VATT and Kuiper features were both 1\% depths, whereas the LOT feature was 0.5\%, and the LOT and Kuiper features were both $\sim40-45$ minutes in duration, whereas the VATT feature was $~\sim60$ minutes (see Figures~\ref{fig:EIC9FL} and ~\ref{fig:EIC9LC}). The weather at the site for all three events was nominal - no clouds during the LOT and VATT events, and light cirrus at the Kuiper in the east as the target was setting. Our observations are done with intra-exposure guiding every few seconds on a bright star near the target, and there were no guiding errors during the observations so the target and reference star centroids remained consistent.

This star was also observed by TESS in Sector 19, and was found to have stellar variability with an amplitude of $\sim$1\% on a 0.99 day period.  Once that trend was removed, a low-confidence recurring event (SNR = 4.9, Signal Detection Efficiency (SDE)$\sim$6.5) was found with a period of 2.883 days and a transit depth of 0.7\%.  Notably, the ephemerides aligned with both the LOT and VATT detections -- 12 orbits of separation with a period of 2.883 days between the two sets of data.  However, additional observations from Lulin and Calar Alto ruled out additional transits during those observation windows, casting doubt on the veracity of the original transit observations.

\begin{figure*}
    \centering
    \includegraphics[width=\linewidth]{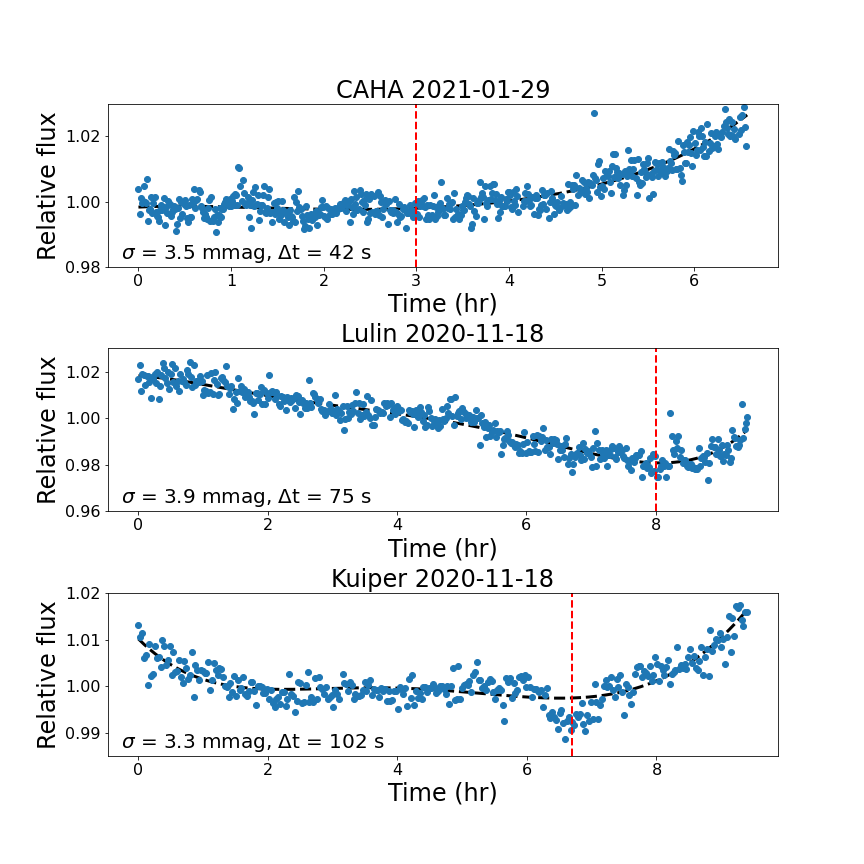}
    \caption{Additionally detrended light curves for EIC 9. Top: Follow-up light curve from Calar Alto in January 2021 with no transit feature near the expected mid-transit time from the TESS period (red dashed line). Middle: LOT light curve from November 2020 with the expected mid-transit time from TESS marked. No transit-like signal is visible. Bottom: Kuiper telescope light curve from November 2020, with the expected mid-transit time from a 2.886-day period. Here, a transit-like feature is visible.}
    \label{fig:EIC9add}
\end{figure*}

Additional analysis of the light curves from the LOT and VATT events found both to be highly dependent on systematics, as the VATT event was at high airmass and the LOT event is cancelled out by the detrending procedure (compare the top panel of Figure~\ref{fig:EIC9FL} with the top panel of Figure~\ref{fig:EIC9LC} and the middle panel of Figure ~\ref{fig:EIC9add}). The event in the Kuiper telescope light curve remained even through additional detrending. Previous observations from the Bok telescope in 2019 weren't sensitive enough to rule out a transit of the given depth, but new data from the Calar Alto $1.23\,\mathrm{m}$ telescope in January 2021 ruled out the $\sim2.883$-day period for the candidate (see Figure~\ref{fig:EIC9add}). Transit fitting of the signal from the Kuiper event put the period at 3.003 days with a 68\% confidence interval of [1.735, 3.781] days. Based on our observing coverage of the target, we likely would have seen multiple transits of any planets within a 5-day period.

Eventually, we determined that our effective phase coverage was providing diminishing returns on continued observations beyond $\sim$40 nights of data, so with no further transit signals we suspended the follow-up of this target and returned to the standard survey monitoring procedures. The event could be a transit, but we would likely have seen an additional transit in our observations unless the data quality was consistently poor. At this point the origin of the detected signal remains unexplained. It may have been caused by telluric contamination such as variable precipitable water vapor~(see discussion in Sect.~\ref{sec:PWV}), which we shall estimate qualitatively here. Typical amounts of water vapor in the Santa Catalina Mountains of \SI{2}{\milli\meter} have been measured \citep{Warner1977}. For a bandpass similar to GG495, this was estimated to cause  a \SI{\sim 3}{\percent} flux difference due to PWV for an M8 target star \citep{Murray2020, Pedersen2023}. According to these data, a PWV variation on the order of \SI{1}{\milli\meter} would be consistent with a \SI{1}{\percent}-level flux change, making PWV variations a potentially viable source of the observed signal.

The \texttt{TLS} results from the full set of light curves of this target confirm our analysis of the potential signal.  There was no strong event matching the period or any multiples of the expected transit from the TESS data, and the best transit model that \texttt{TLS} found had too short of a duration to be a physical transit of an exoplanet.  In addition, none of the best-fit models from any of the precision cuts included the Kuiper event as a possible transit match.

\subsection{Light curve analysis}

Our observations were collected with seven telescopes located on three different continents. We developed a uniform observing protocol and data reduction and analysis approach that provided uniform final products, where differences are primarily due to the sensitivity of the instrument (a combination of the telescopes' light-collecting area and the quantum efficiency of the detectors).  Our data reduction and analysis approach was developed to simultaneously maximize sensitivity (finding many possible transit candidates and following up on them) and efficiency (observing many targets where follow-up monitoring requires a substantial amount of telescope time).

The EDEN survey's sensitivity is demonstrated by the injection-and-retrieval tests we performed (see Section~\ref{subsec:sense}). Its efficiency is, perhaps counter-intuitively, reflected in the overall scarcity of possible detections: our sensitivity analysis shows that our observations can very efficiently detect short-period planets (P $<$1.5~d), even if they are relatively small (R$_p<1.5 R_\oplus$).  Yet, during our survey of over 2,450 hours, we detected only one target with a potential single transit (see Section~\ref{subsec:J0419+4233}). The combination of high sensitivity and the very low number of false positives demonstrates high efficiency and reliability, as well as a robust approach for data reduction and analysis, which are particularly important for any targeted survey. 
In order to support future surveys, we commit to making the complete EDEN data analysis package available on a collaborative basis.

\subsection{No planets detected}

Project EDEN survey's extensive dataset yielded multiple possible detections, but only EIC 9 was not identified immediately as a false positive. Further observations successfully eliminated it as a potential planet candidate, leaving no detected transit in our survey. Translating non-detections into constraints on the exoplanet demographics and occurrence rates is non-trivial, and requires a careful sensitivity analysis (see, e.g., \citealt[][]{Obermeier2016,Kunimoto2020,Sagear2020}). 

We performed injection-and-recovery based Monte Carlo sensitivity assessments for our targets and datasets, which mimic signals which would have been introduced by planets. In the initial EDEN survey paper \citep[][]{Gibbs2020}, where our method was initially introduced, the efficiency was compared to (blind) injection and recovery by four team members. We found that the human team's ability to identify randomly injected planets is comparable to the algorithmic identification. Therefore, the automatic injection-and-recovery sensitivity maps presented in Appendix~\ref{sec:maps} can be considered as a highly reliable assessment of the sensitivity of our dataset for each planet as a function of planet period and radius.  The non-detection, in combination with the carefully characterized sensitivity limits, provides an important opportunity to place constraints on the occurrence rates of exoplanets around late M~dwarfs.

\subsection{Occurrence rates for different categories of planets}
\label{subsec:DiscussionTRAPPIST1b}

While we obtain a sensitivity of $\sim30\%$ to transiting TRAPPIST-1b-like planets, the completeness with regard to these planets is low due to the small transit probabilities. Our Monte Carlo assessment of 10,000 mock surveys showed that even if every star had a TRAPPIST-1b-like planet or TRAPPIST-1 planetary system analog, we would detect planets either \SI{15}{\percent} or \SI{21}{\percent} of the time.  In addition, \citet{Burn2021} found in planet formation models that occurrence rates of planets of all sizes drop from early M~dwarfs to late M~dwarfs. Based on a pebble accretion formation model, \citet{Mulders2021} suggest a drop in rocky planets toward the substellar boundary.

We can compare our results to other measurements from ultra-cool dwarf (UCD) surveys.  The full TRAPPIST survey analysis \citep[][]{Lienhard2020} found that their expected number of detections, given every UCD had a TRAPPIST-1~b-like planet in the system, is 0.52.  Thus, the detection of TRAPPIST-1~b in the survey sample of 40 UCDs implies a lower limit for the occurrence rate of TRAPPIST-1~b-like planets of 10\% at 95\% confidence.  \citet[][]{SestovicDemory2020} examined the K2 light curves of 702 UCDs and were able to rule out TRAPPIST-1-like planetary systems around every UCD with 96\% confidence, while \citet[][]{Sagear2020} examined 827 K2 UCD light curves and found that the occurrence rate of TRAPPIST-1b analogs was limited to $<57\%$. The EDEN survey, with roughly half the number of targets as the full TRAPPIST survey analysis, shows a similar expected number of detections. Thus, even a targeted EDEN-like survey expanded out to >100 UCDs would still have at least a 50\% chance of detecting no planets. 



The case of giant planets is a separate consideration. Compared to FGK stars, M~dwarfs host fewer giant planets \citep{Endl2006,Johnson2010,Bonfils2013,Sabotta2021}, which is in line with predictions of current planet formation models \citep[e.g.,][]{Miguel2016,Liu2019,Mulders2021,Burn2021}.  The giant planet occurrence appears to decline linearly with decreasing stellar mass~\citep{Johnson2010,Ghezzi2018}, although the continuation of this trend into the less explored late M~dwarf regime has recently been questioned~\citep{Jordan2022,Schlecker2022}. Under the optimistic assumption of a short-period ($P\sim1$ day) giant planet around every star, \SI{27}{\percent} of simulated EDEN surveys would still yield zero detections. Our data thus allows us to reject the hypothesis of one giant planet per star at a \SI{73}{\percent} significance level. For ultra-short-period giant planets ($P=0.5$ days), our constraint on the occurrence rate at 95\% confidence is $f_\mathrm{large} <$ \SI{75}{\percent}. We conclude that our non-detection is consistent with previous giant planet occurrence rates while not ruling out a moderate increase around late M~dwarfs as recently suggested by \citet{Schlecker2022}. A larger sample will be needed to provide stronger constraints.  Assuming a comparable completeness for an extension of the survey, we find that a sample with at least $\sim 50$ stars is needed to exclude the hypothesis that every star hosts a (hot) giant planet with \SI{95}{\percent} confidence.

\section{Summary} \label{sec:summary}

The occurrence rate of TRAPPIST-like planets is not well constrained, as well as the occurrence rate of small planets around earlier M~dwarfs and its dependency on spectral subtype \citep[e.g.,][]{Hardegree-Ullman2019,Mulders2021}.  We have performed a northern-visible volume-complete survey out to \EDENVolume{} of the late (M7--L0) single and quiescent northern M~dwarfs searching for transiting planets to test this hypothesis.  The key findings of our study are as follows:

\begin{itemize}
    \item We observed $>$ \avgtime{} for 22 target stars observable for $>$ 3 hours per night from the northern hemisphere and present high-precision light curves and data sets for all targets.

    \item We provided an improved planet-injection-based sensitivity analysis to remove low-sample-size noise.

    
    \item We carefully analyzed the data and found no convincing periodic planetary transit signals that held up to advanced scrutiny.  We found one possible candidate in EDEN and TESS data, but further observations and analysis ruled out the origin of the signals as a repeating transit feature.
    
    \item The combination of high sensitivity and very low number of false positives demonstrated the high efficiency of our data analysis and planet identification process.
    
    
    \item We tested the fraction $f_\mathrm{Trb}$ of late M~dwarfs hosting a TRAPPIST-1~b analog planet and $f_\mathrm{Tr}$ of a TRAPPIST-1 analog system. Given no detections in our survey, and the well-characterized sensitivity, we cannot exclude $f_\mathrm{Trb} < 100\%$ (15\% of surveys would detect a planet) nor an $f_\mathrm{Tr} < 100\%$  (21\% of surveys would detect a planet).  
    

    \item Additionally, we tested the fraction $f_\mathrm{large}$ of late M~dwarfs hosting a giant planet at a short orbital period $P=1.05$ days. 
    We found that we cannot exclude $f_\mathrm{large} < 100\%$ with 73\% of simulated surveys detecting a planet. However, at 0.5 day orbital periods, we can constrain $f_\mathrm{large} < 75\%$ at a 95\% confidence level.
    
    
\end{itemize}

Our EDEN observations provide the most sensitive volume-complete photometric monitoring of late M~dwarf stars to date and upper limits on the short-period planet population around TRAPPIST-1-like hosts. The observations presented here can guide future studies of the targeted systems and be used to test models of planet formation and evolution around the smallest stars.\\

The authors would like to thank N\'estor Espinoza and Jos\'e P\'erez Angel Ch\'avez for major contributions to the EDEN data processing pipeline, Allie Mousseau for creating the EDEN target catalog, Quentin J. Socia for observations at the Kuiper 61" and the VATT 1.8 m telescopes, and Roberto Gualandi for his technical assistance at the Cassini telescope. The results reported herein benefited from collaborations and/or information exchange  within NASA's Nexus for Exoplanet System Science (NExSS) research coordination network sponsored by NASAs Science Mission Directorate. This material is based upon work supported by the National Aeronautics and Space Administration under Agreements No. NNX15AD94G (Earths in Other Solar Systems, PI: Apai) and No. 80NSSC21K0593 (Alien Earths, PI: Apai). This research has made use of the NASA Exoplanet Archive, which is operated by the California Institute of Technology, under contract with the National Aeronautics and Space Administration under the Exoplanet Exploration Program. This research made use of Photutils, an Astropy package for detection and photometry of astronomical sources \citep[][]{Bradley2019}. This research made use of the Cassini 1.52 m telescope, which is operated by INAF-OAS ``Osservatorio di Astrofisica e Scienza dello Spazio'' of Bologna in Loiano, Italy. This publication has made use of data collected at Lulin Observatory, partly supported by MoST grant 109-2112-M-008-001. We thank the mountain operations staff at University of Arizona, Mt. Lemmon Sky Center, Lulin Observatory, Calar Alto Observatory, Loiano Telescopes, Mt. Graham International Observatory, Vatican Advanced Technology Telescope, and Kitt Peak National Observatory. BVR thanks the Heising-Simons Foundation for support. LM acknowledges support from the ``Fondi di Ricerca Scientifica d'Ateneo 2021'' of the University of Rome ``Tor Vergata''. TND acknowledges support provided by the Alexander von Humboldt Foundation in the framework of the Sofja Kovalevskaja Award endowed by the Federal Ministry of Education and Research.  An allocation of computer time from the UA Research Computing High Performance Computing (HPC) at the University of Arizona is gratefully acknowledged.

\begin{large}\textit{Author contributions:}\end{large}
JD coordinated the observations of the EDEN target list, performed the sensitivity analysis, and drafted the manuscript. DA, TH, PG, and W-PC are the Co-PIs who founded EDEN, defined its scope, goals, and methods, led proposals to fund the project and led the research groups performing the observations at the University of Arizona, the Max Planck Institute for Astronomy, the Vatican Observatory, and National Central University in Taiwan. LM led an additional major component out of Italy and the University of Rome ``Tor Vergata''. DA, KKH-U, MS, KM, and NK provided important assistance for the light curve processing and sensitivity analysis and made major contributions to the manuscript as part of the EDEN ``Tiger Team''. KKH-U assisted in EDEN Survey and Follow-up Observation Program target selection. DA, AG, BVR, and AB created the EDEN data processing pipeline and assisted with the sensitivity analysis. PG, RPB, AB, AG, JD, and MS observed with the VATT 1.8 m telescope. AB, AG, and JD observed with the Bok 2.3 m telescope. AB, JD, KKH-U, and MS observed with the Kuiper 61" telescope. MS and NK coordinated the observing program at MPIA. MS, KM, NK, SB-S, Re.Bu., TND, LF, RF, GP, SS, and JS observed with the Calar Alto 1.23 m telescope. LM and IB observed with the Loiano Cassini 152 cm telescope. C-CN and A-LT coordinated the observing program at NCU. H-CL, C-SL, H-YH, and W-JH observed with the Lulin One-meter Telescope. W-HI and C-LL provided Lulin light curve analysis. Ri.Ba. processed light curves and assisted with Kuiper 61" observations.

%

\facilities{Bok 2.3 m Telescope, Calar Alto Observatory 1.23 m Telescope, Loiano Cassini 152 cm Telescope, Mt. Bigelow Kuiper 61" Telescope, Lulin Observatory Lulin One-meter Telescope (LOT) 1 m, Mt. Lemmon SkyCenter Schulman 32" Telescope, Vatican Advanced Technology Telescope (VATT) 1.8 m, NASA Exoplanet Archive}


\software{astrometry.net \citep[][]{Lang2010}, Astropy \citep[][]{PriceWhelan2018}, \texttt{batman} \citep[][]{Kreidberg2015}, NumPy \citep[][]{Harris2020}, Photutils \citep[][]{Bradley2019}, PyMultiNest \citep[][]{Buchner2014}, SciPy \citep[][]{Virtanen2020}, \texttt{TLS} \citep[][]{Hippke2019a}}, \texttt{W\={o}tan} \citep[][]{Hippke2019b}



\appendix

\section{Sensitivity Maps} \label{sec:maps}

\begin{figure}
    \centering
    \includegraphics[width=0.46\linewidth]{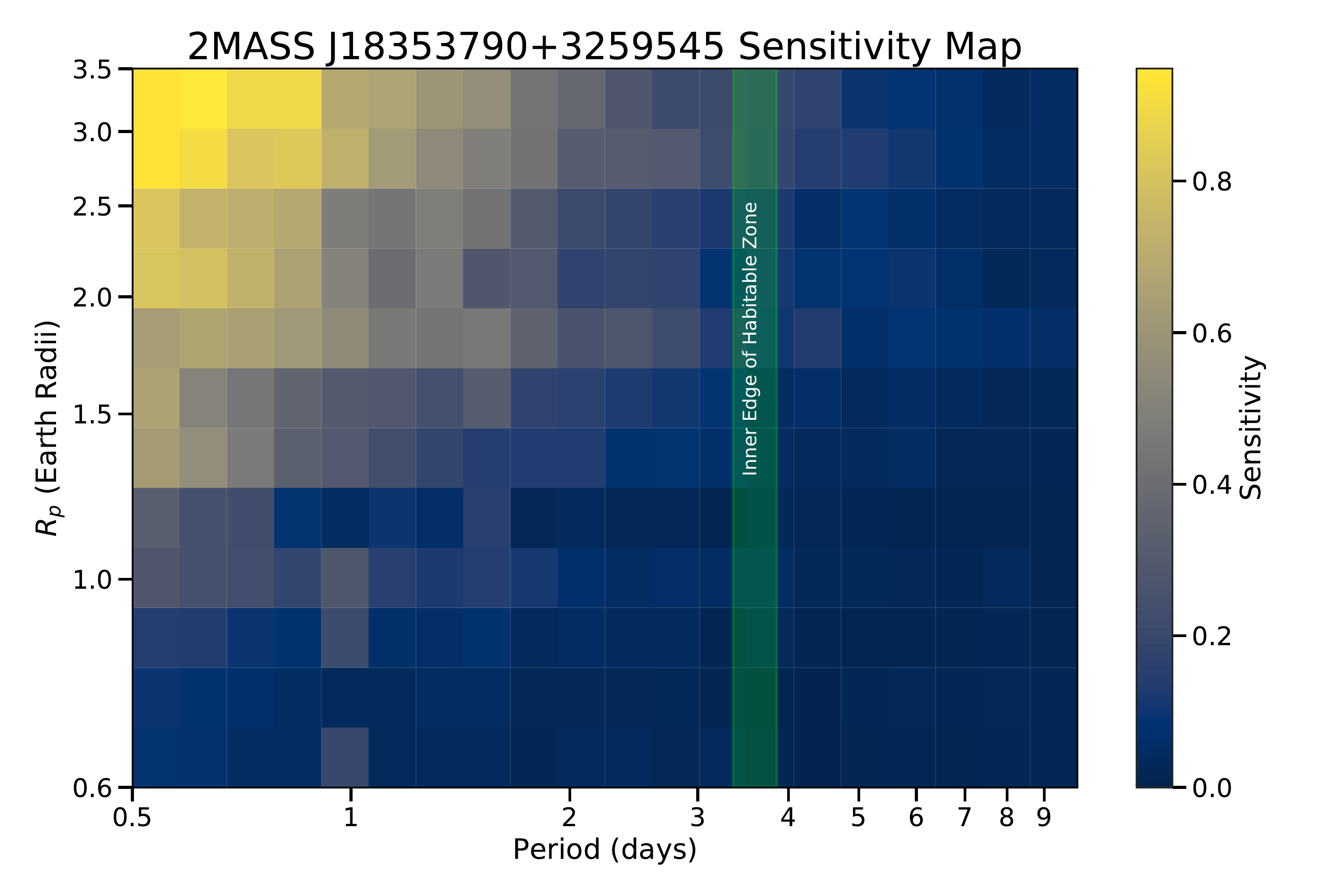}
    \includegraphics[width=0.46\linewidth]{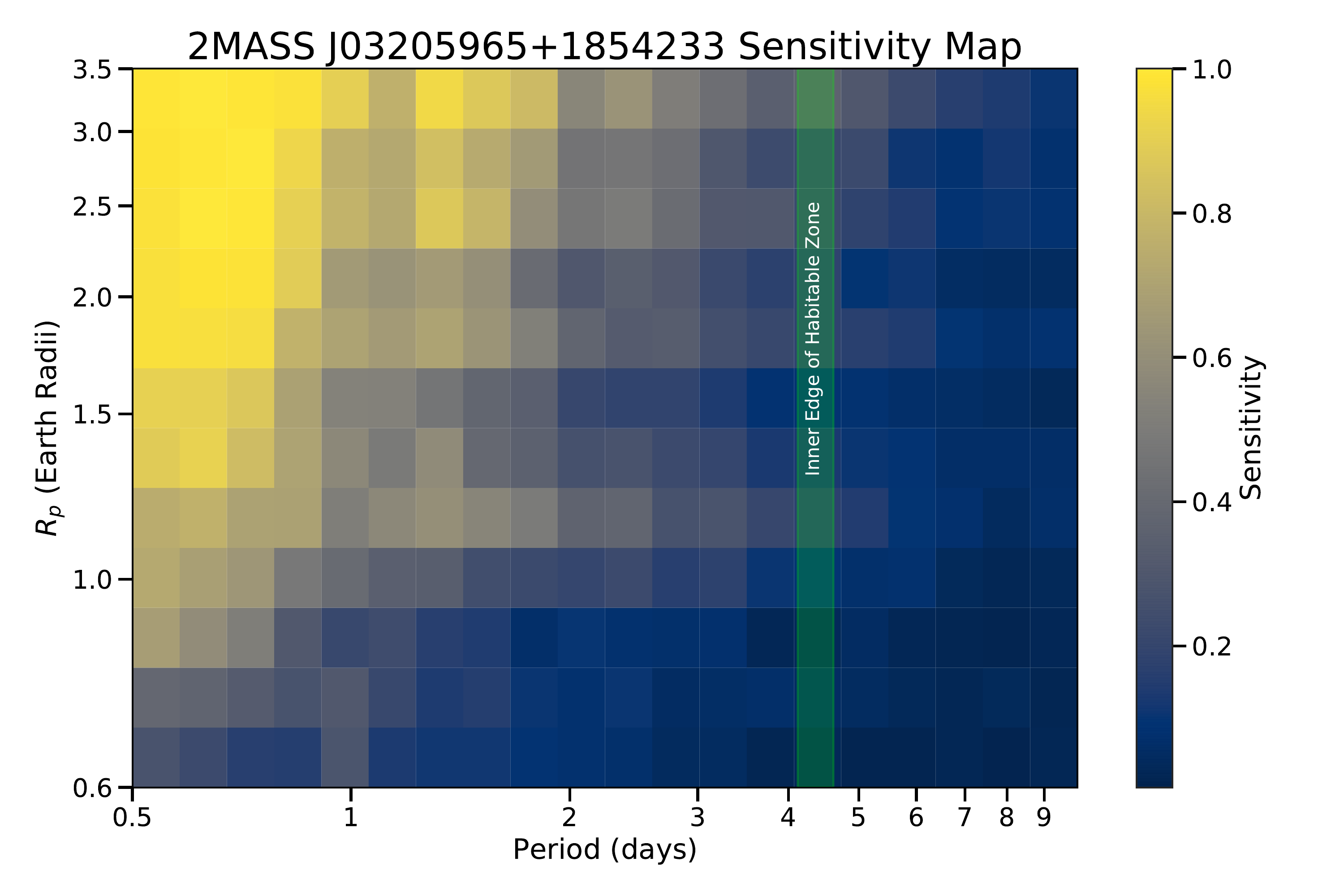}
    \includegraphics[width=0.46\linewidth]{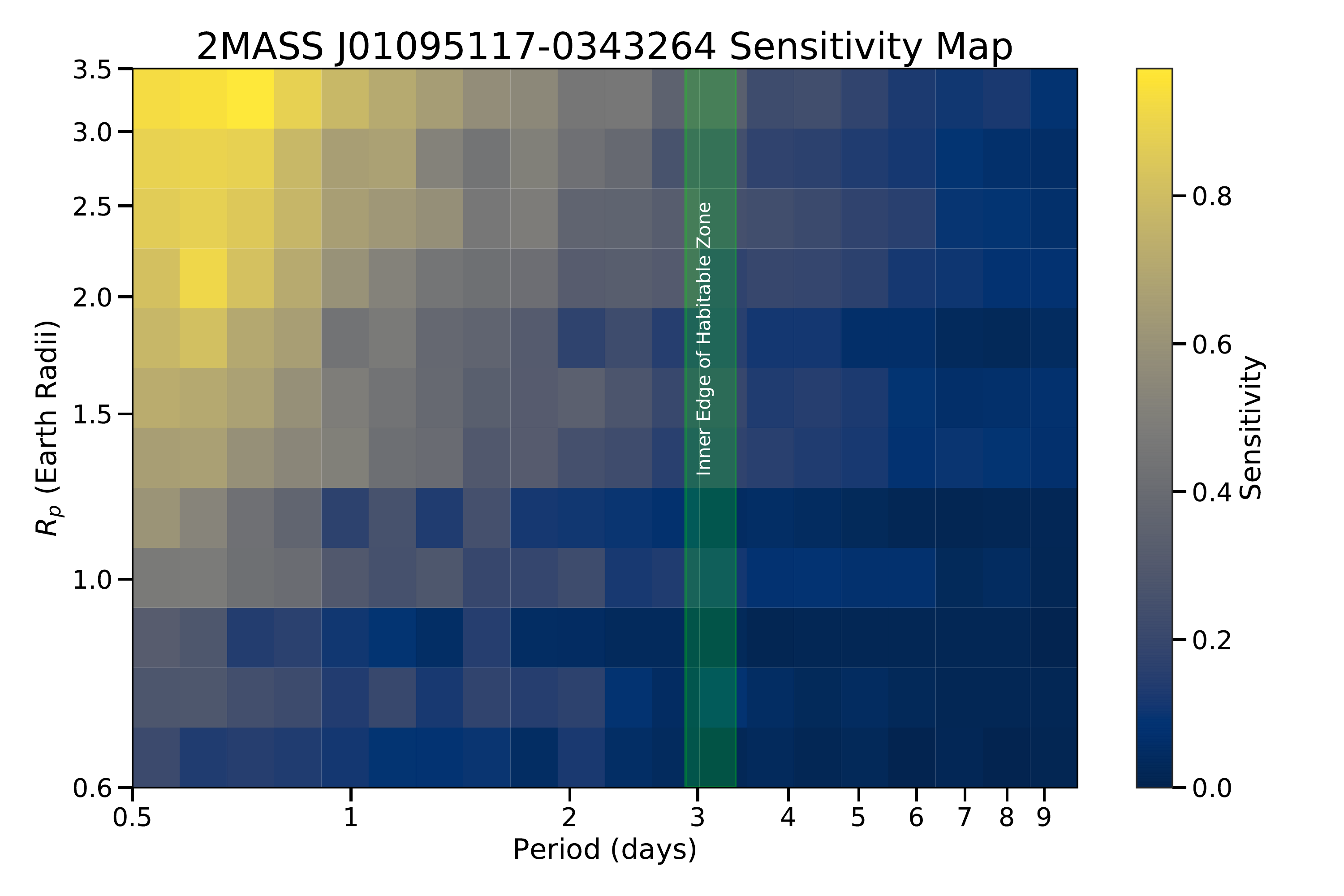}    \includegraphics[width=0.46\linewidth]{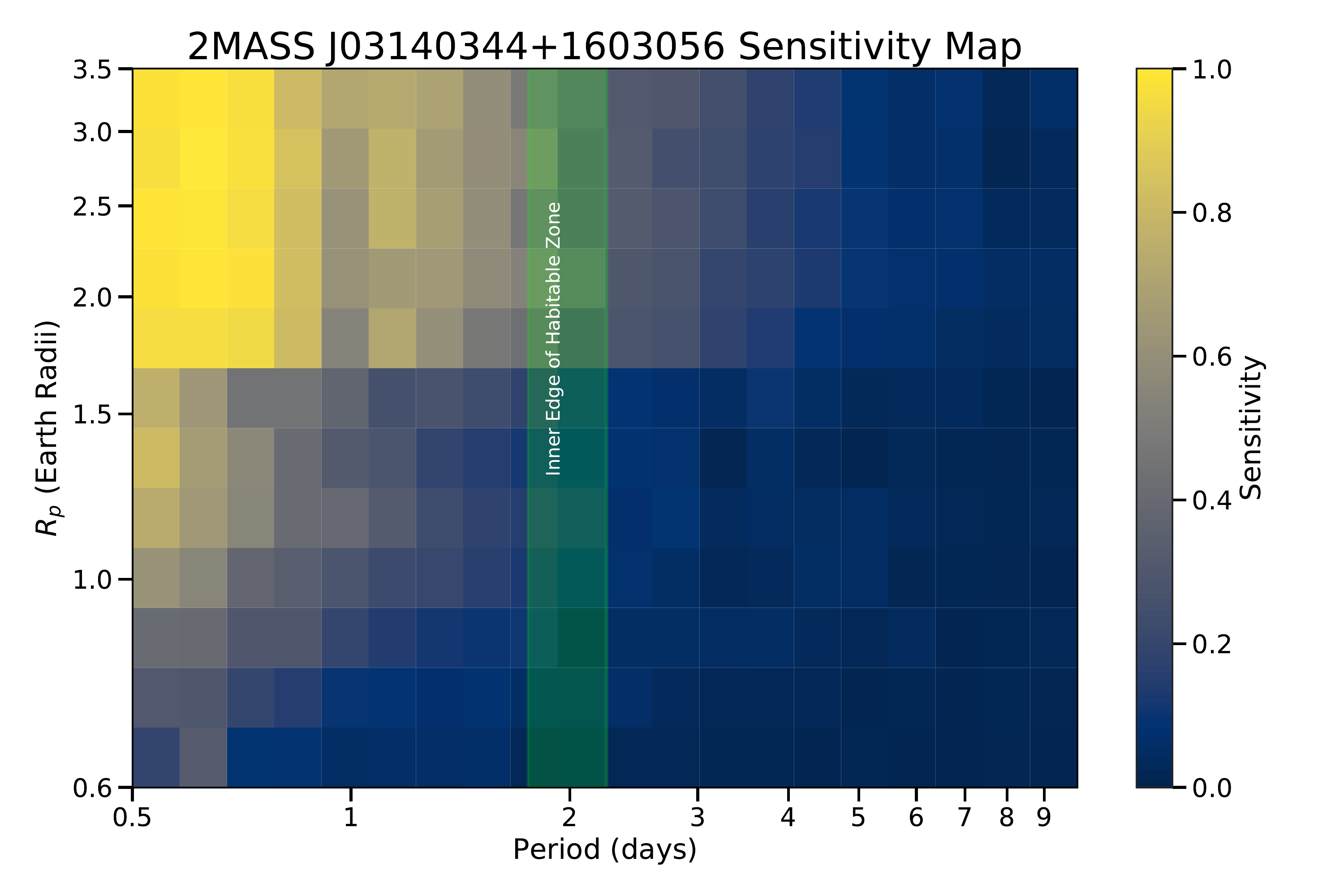}
    \includegraphics[width=0.46\linewidth]{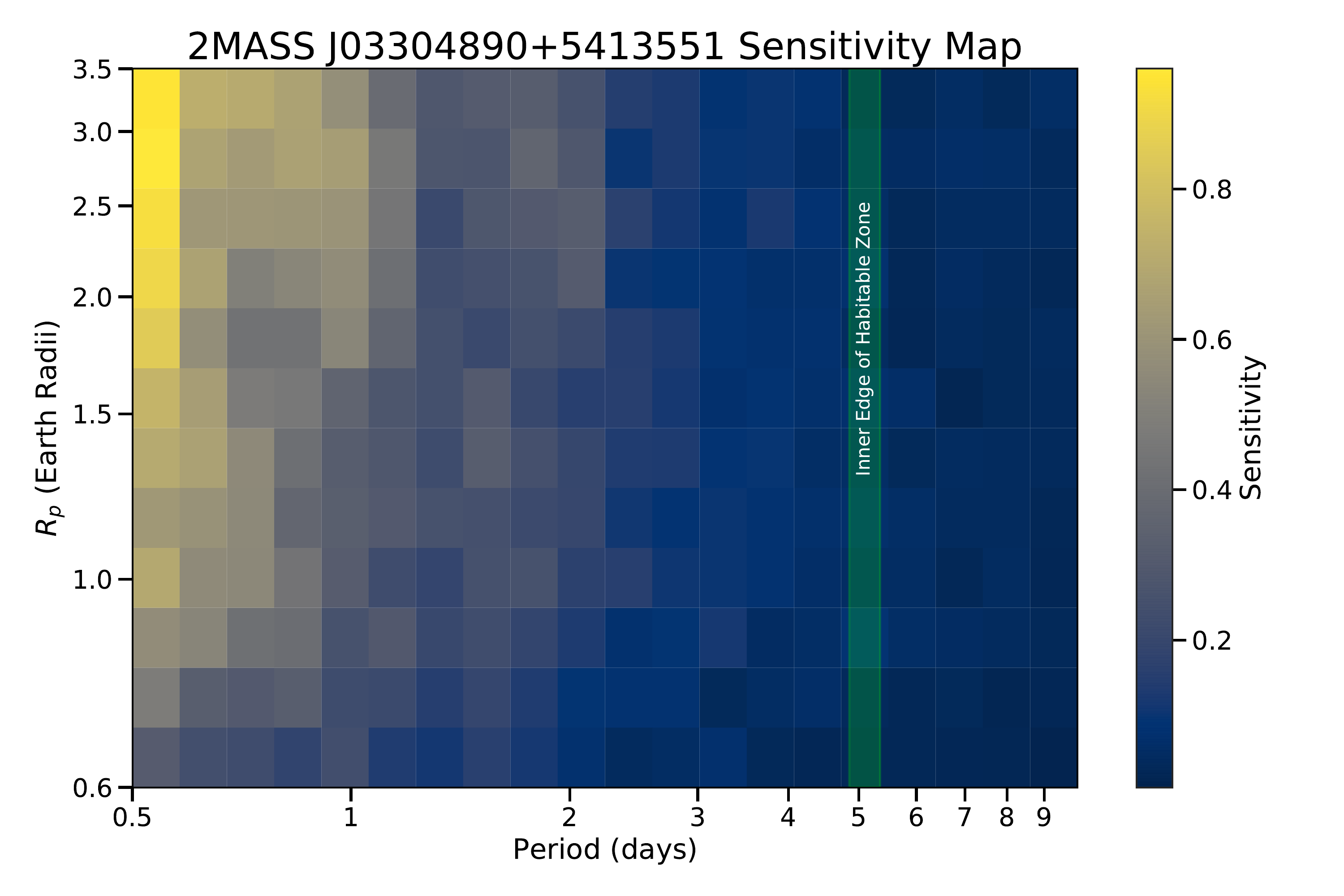}
    \includegraphics[width=0.46\linewidth]{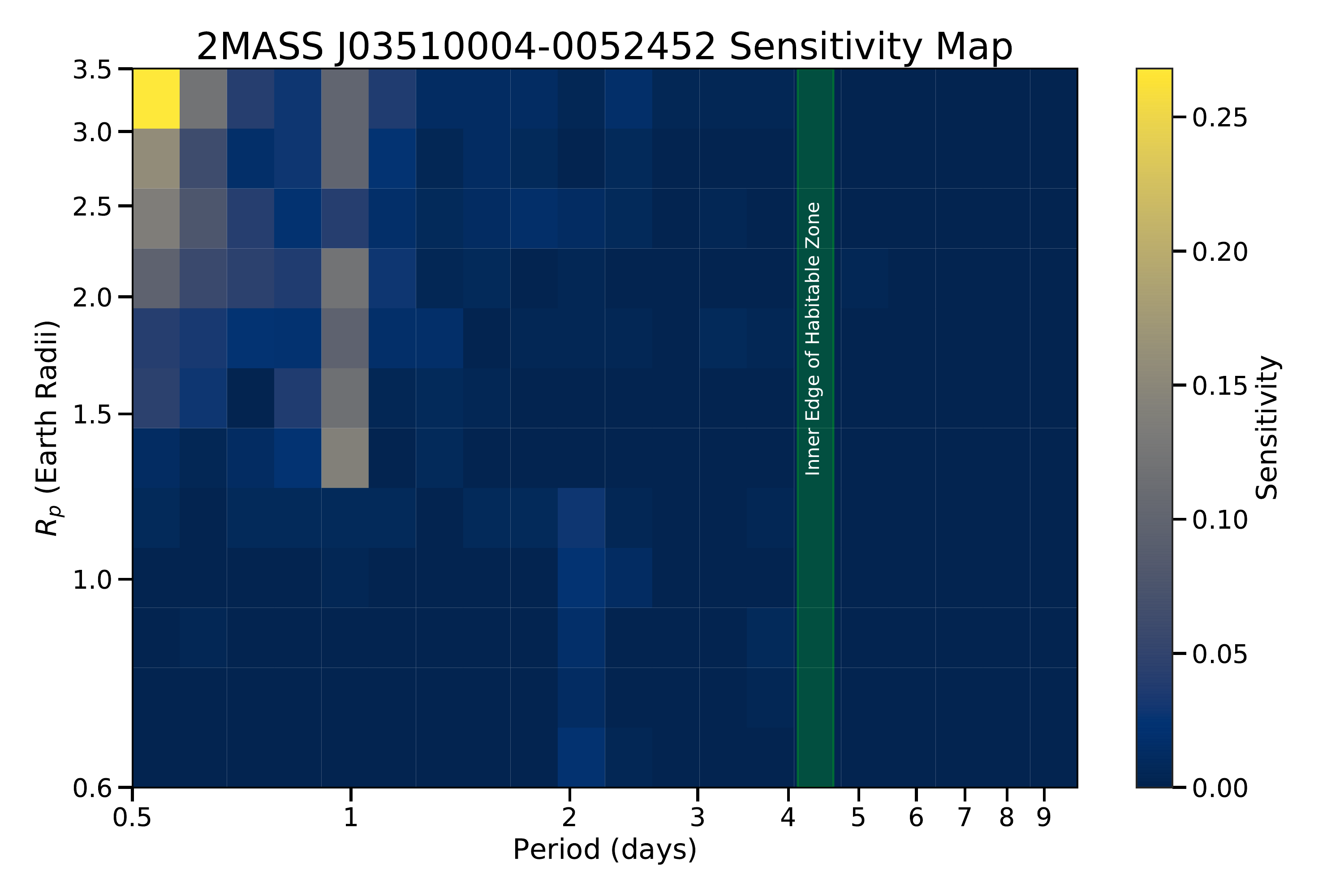}
    \includegraphics[width=0.46\linewidth]{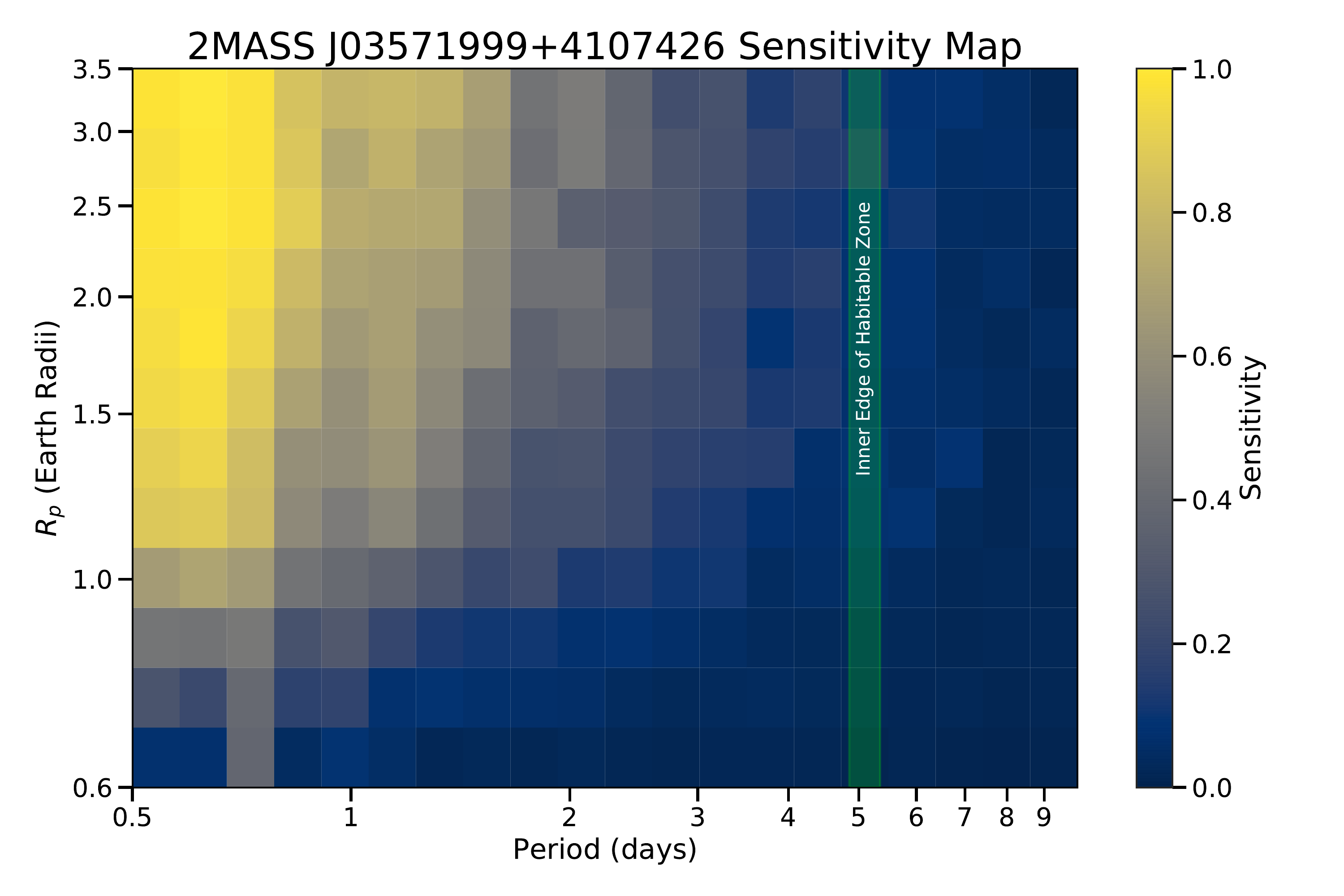}
    \includegraphics[width=0.46\linewidth]{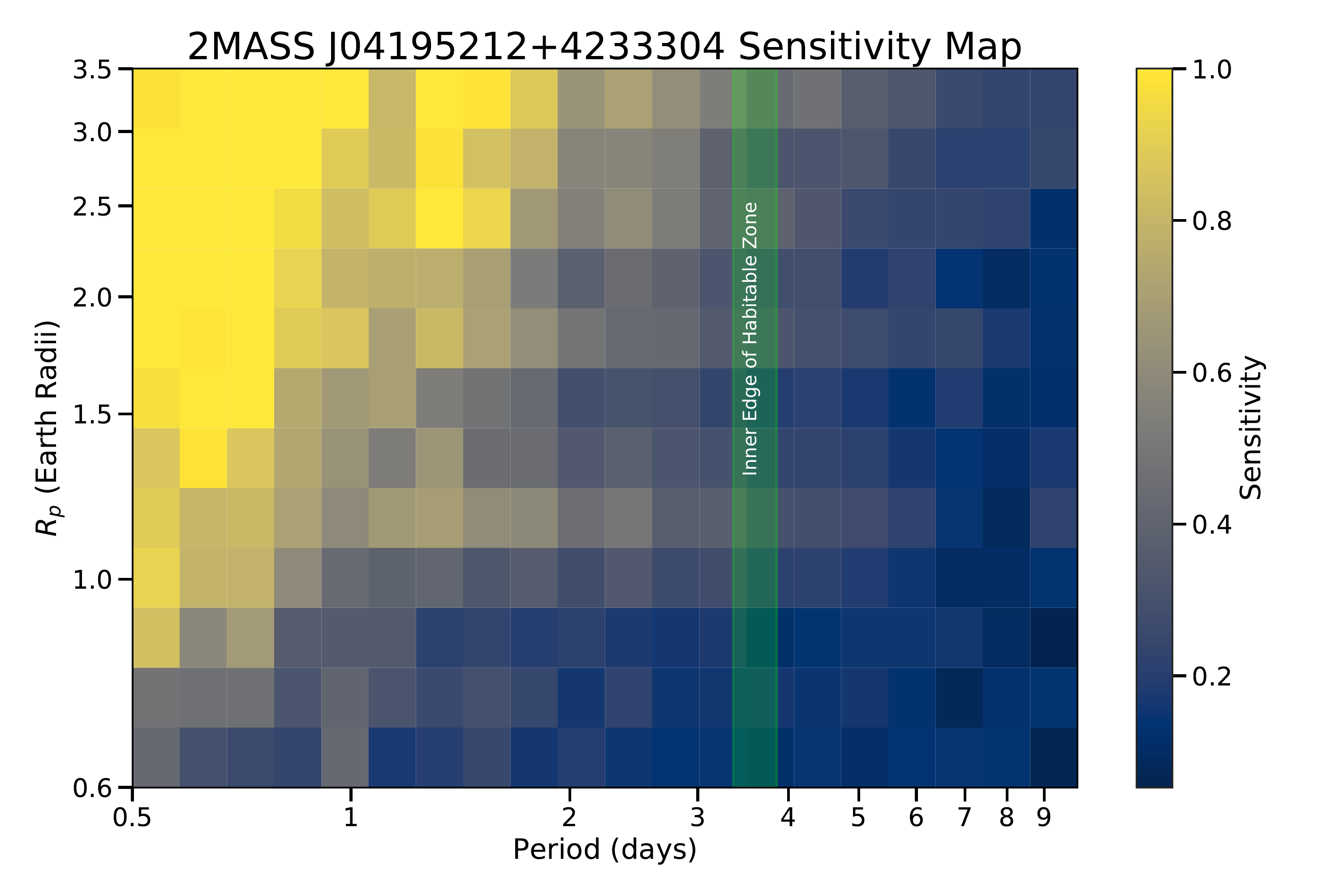}
    \caption{Sensitivity maps for EIC 1, 2, 4, 5, 6, 7, 8, 9}
    \label{fig:sense1}
\end{figure}

\begin{figure}
    \centering
    \includegraphics[width=0.46\linewidth]{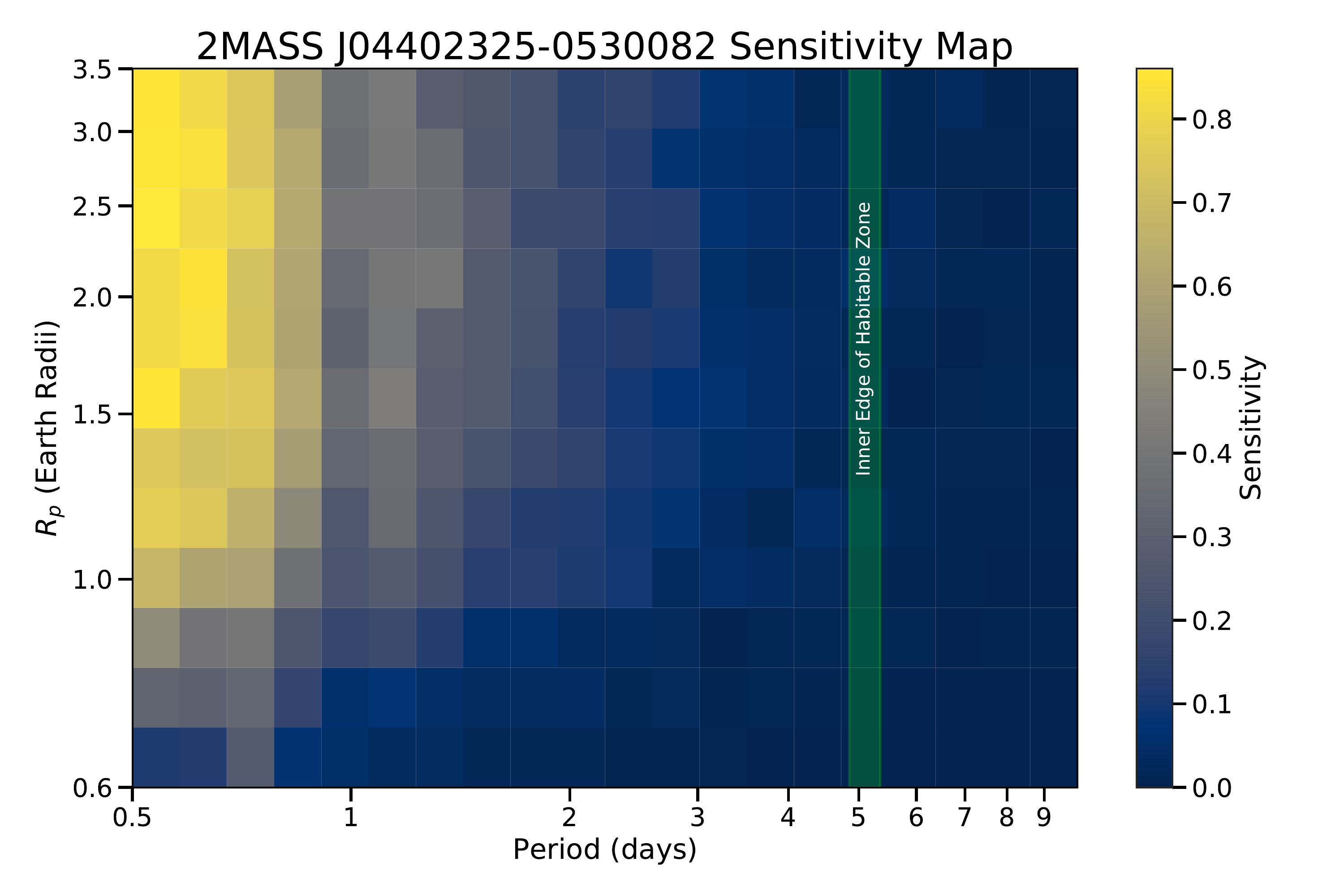}
    \includegraphics[width=0.46\linewidth]{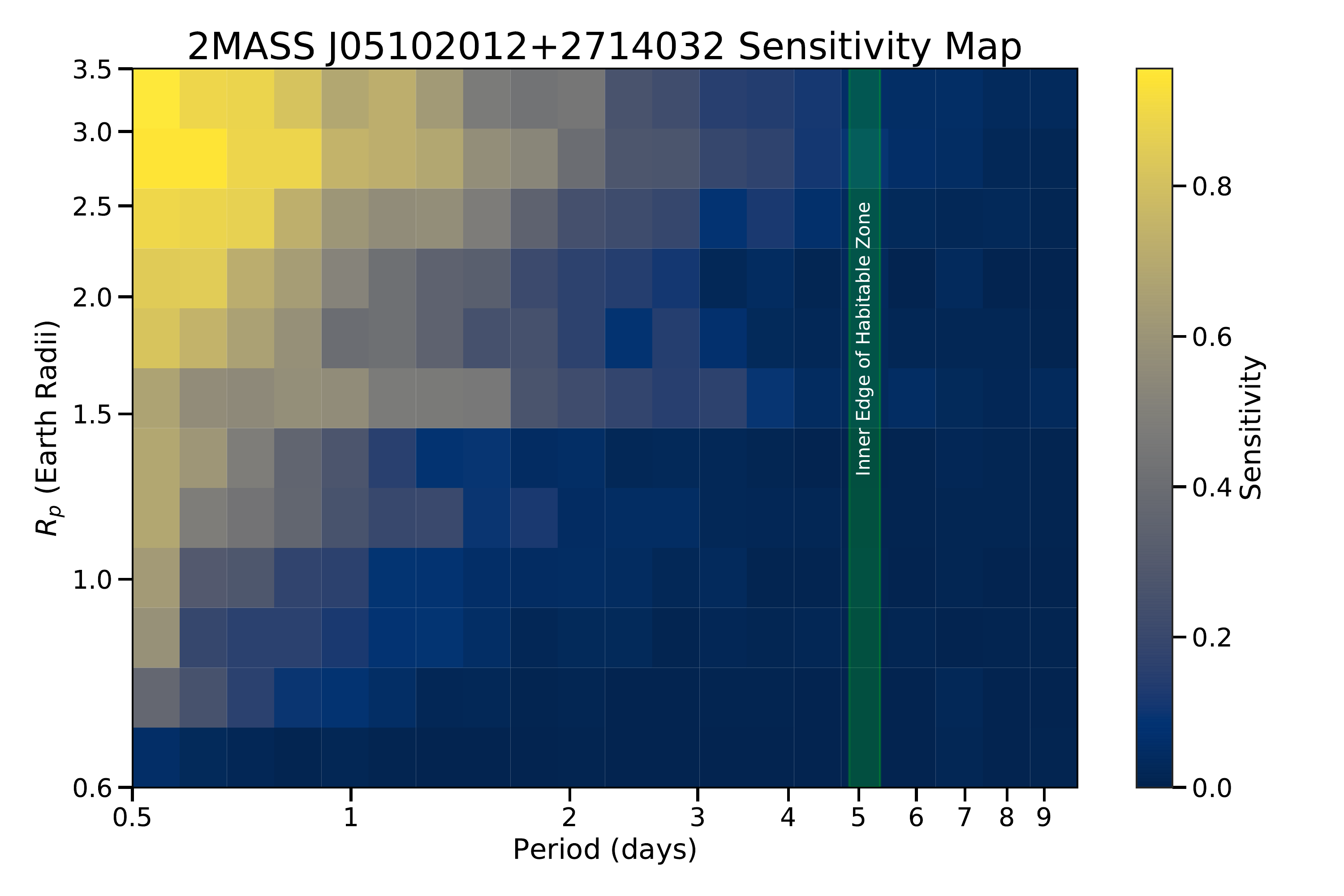}
    \includegraphics[width=0.46\linewidth]{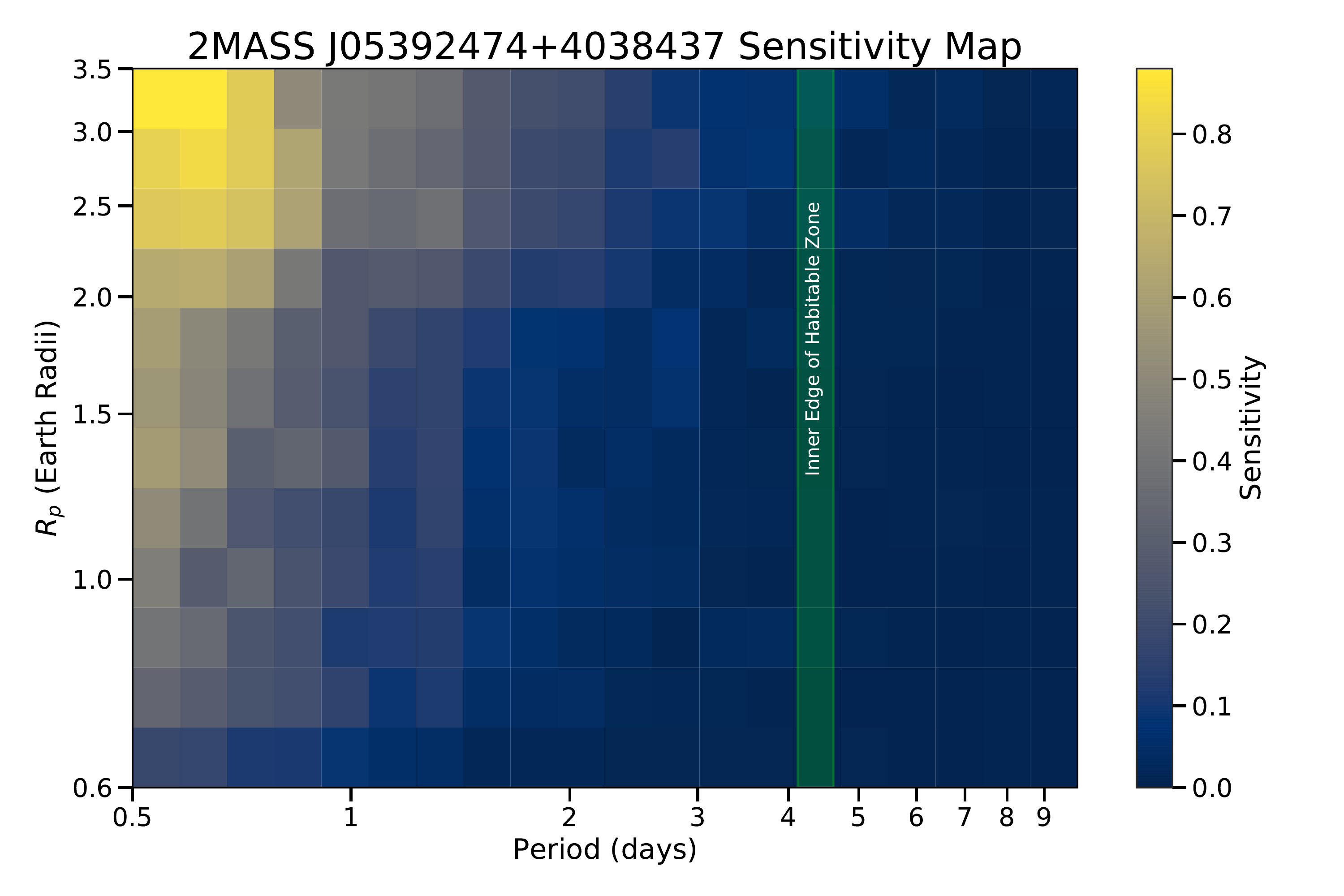}
    \includegraphics[width=0.46\linewidth]{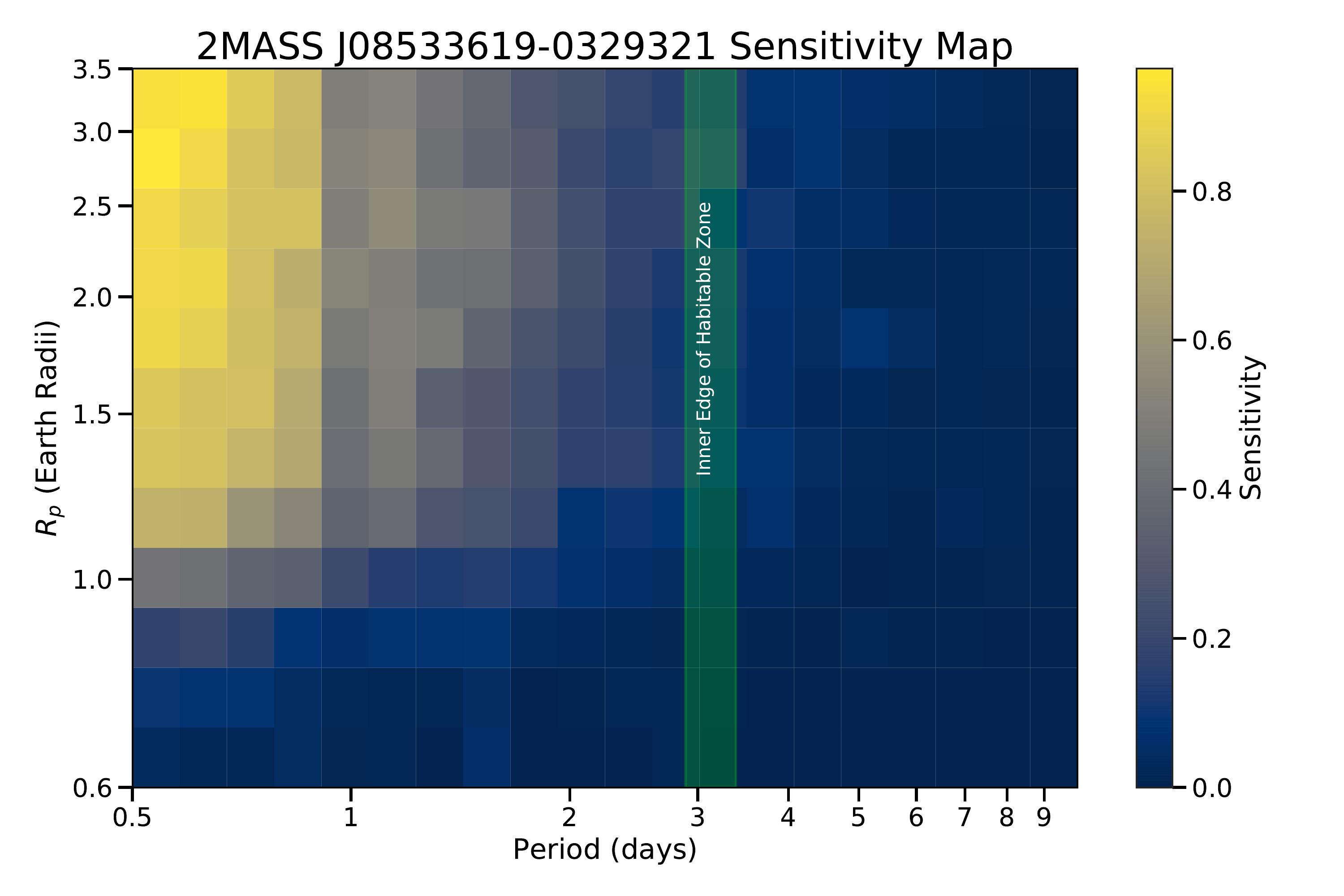}
    \includegraphics[width=0.46\linewidth]{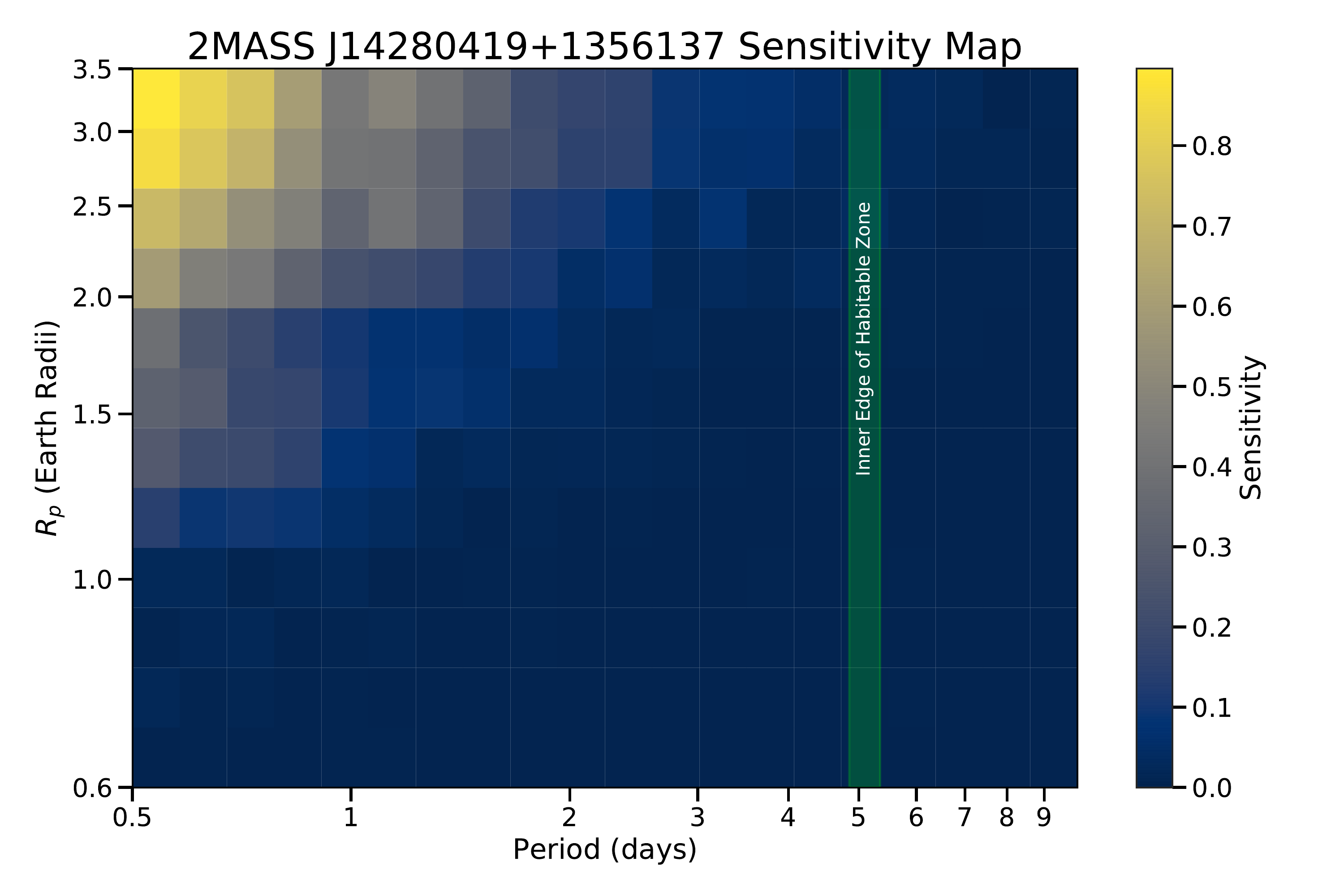}
    \includegraphics[width=0.46\linewidth]{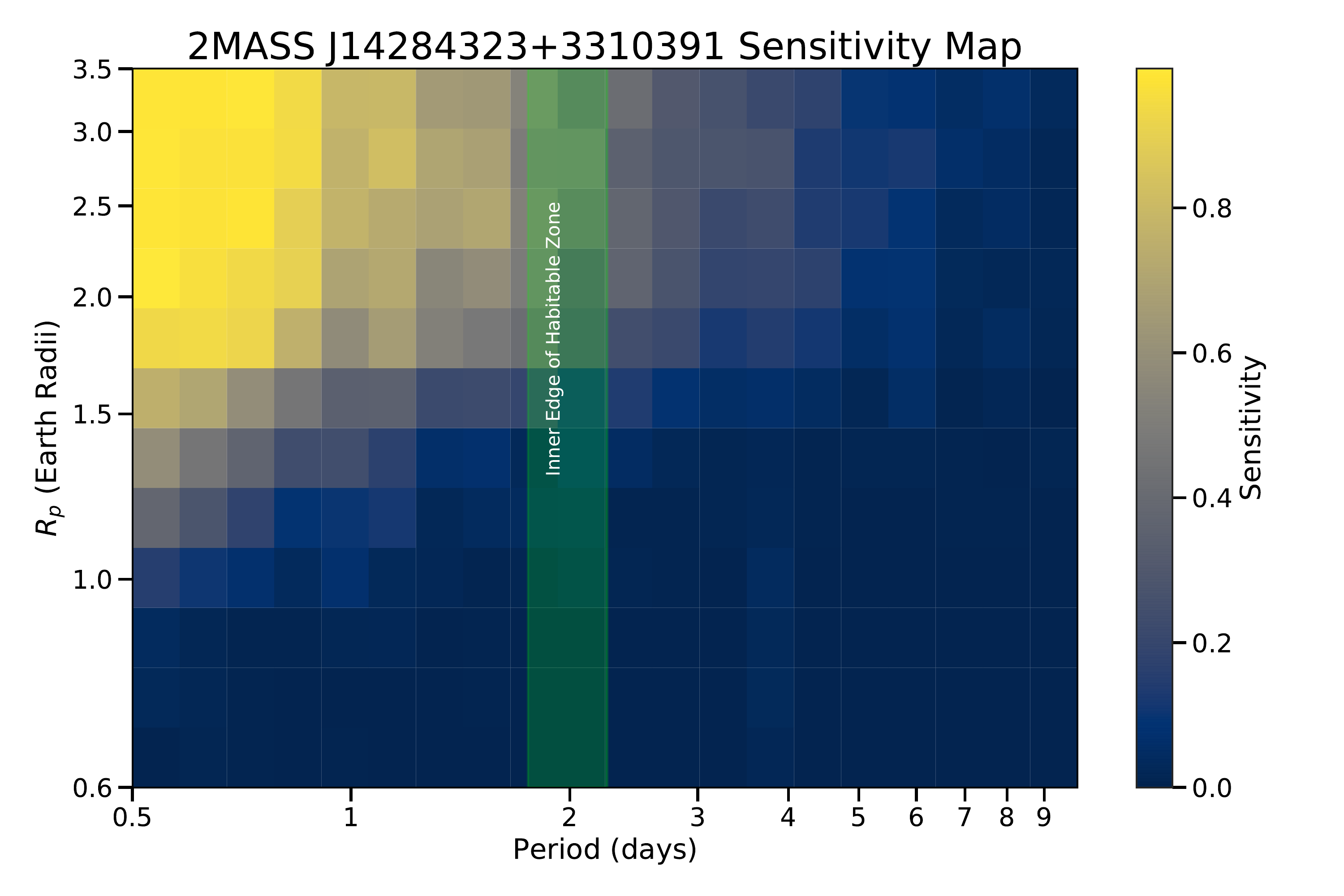}
    \includegraphics[width=0.46\linewidth]{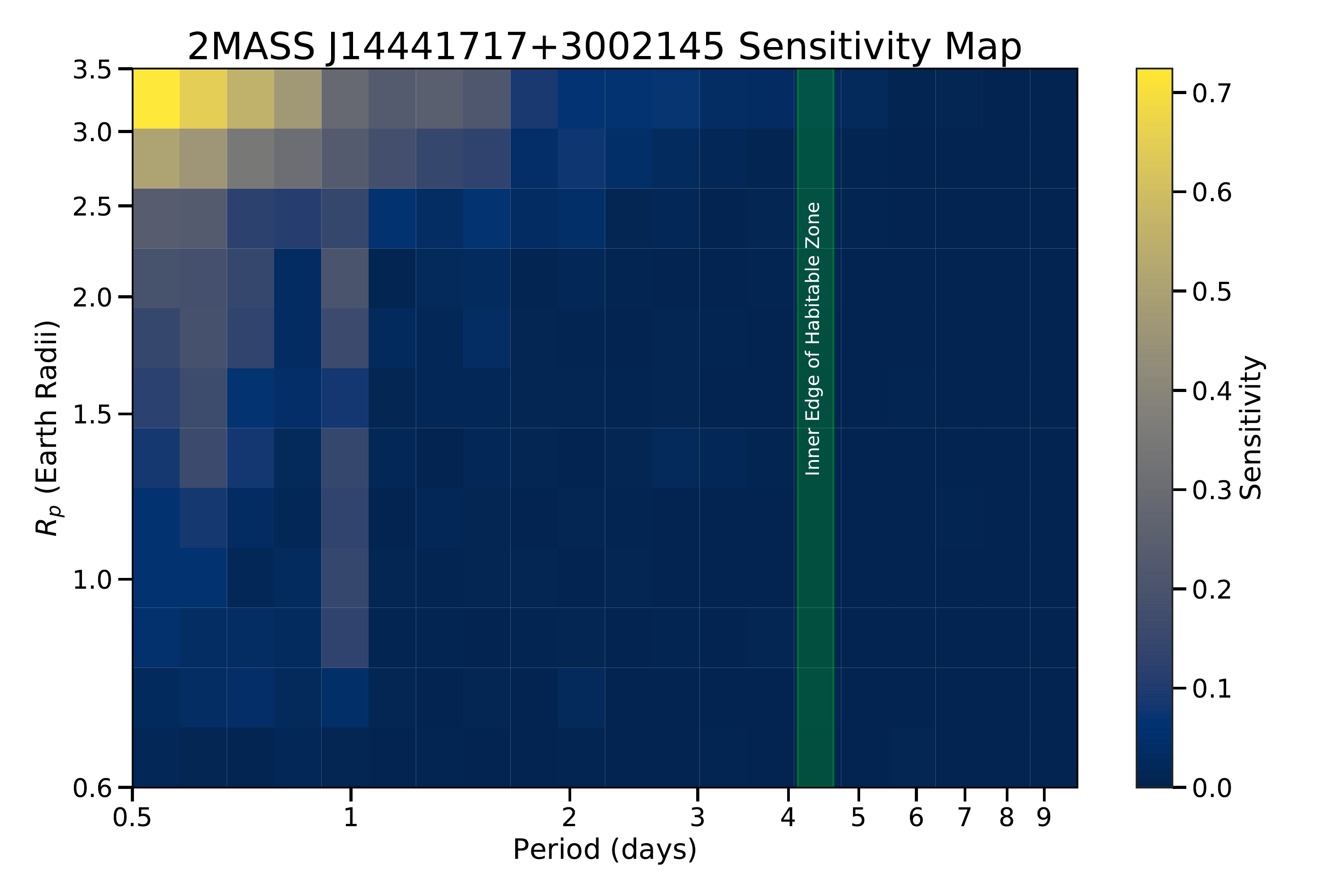}
    \includegraphics[width=0.46\linewidth]{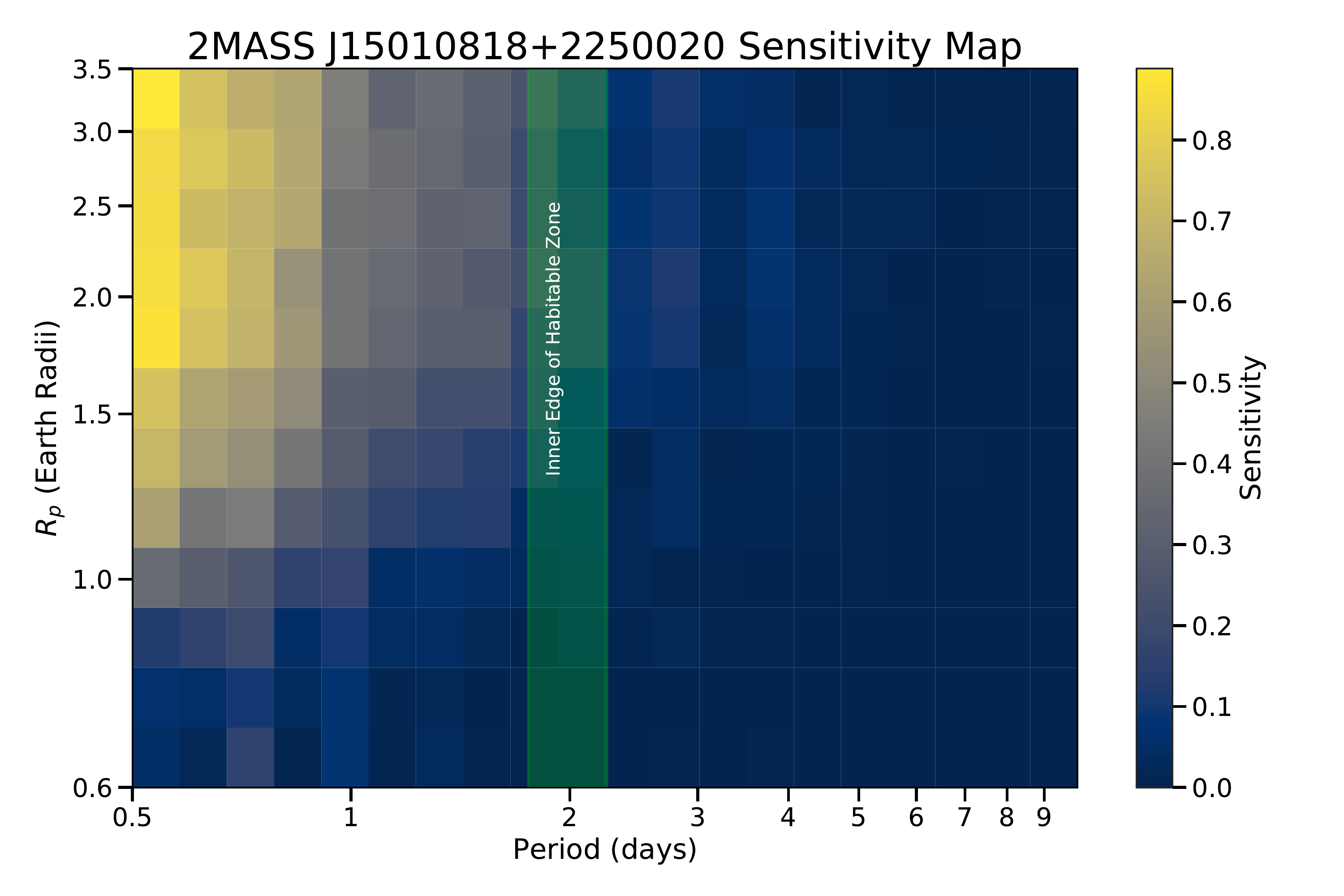}
    \caption{Sensitivity maps for EIC 11, 12, 13, 15, 16, 17, 18, 19}
    \label{fig:sense2}
\end{figure}

\begin{figure}
    \centering
    \includegraphics[width=0.49\linewidth]{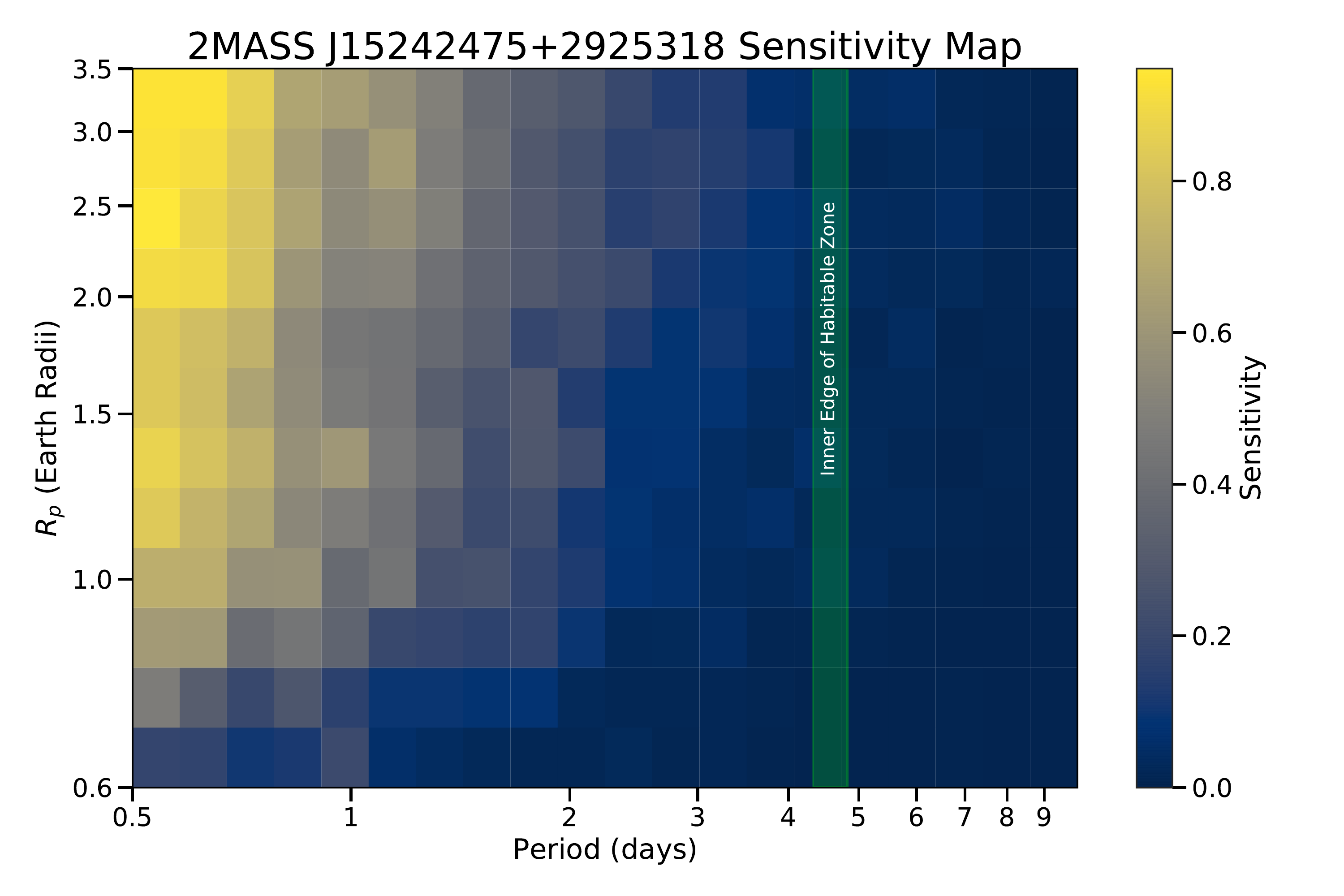}
    \includegraphics[width=0.49\linewidth]{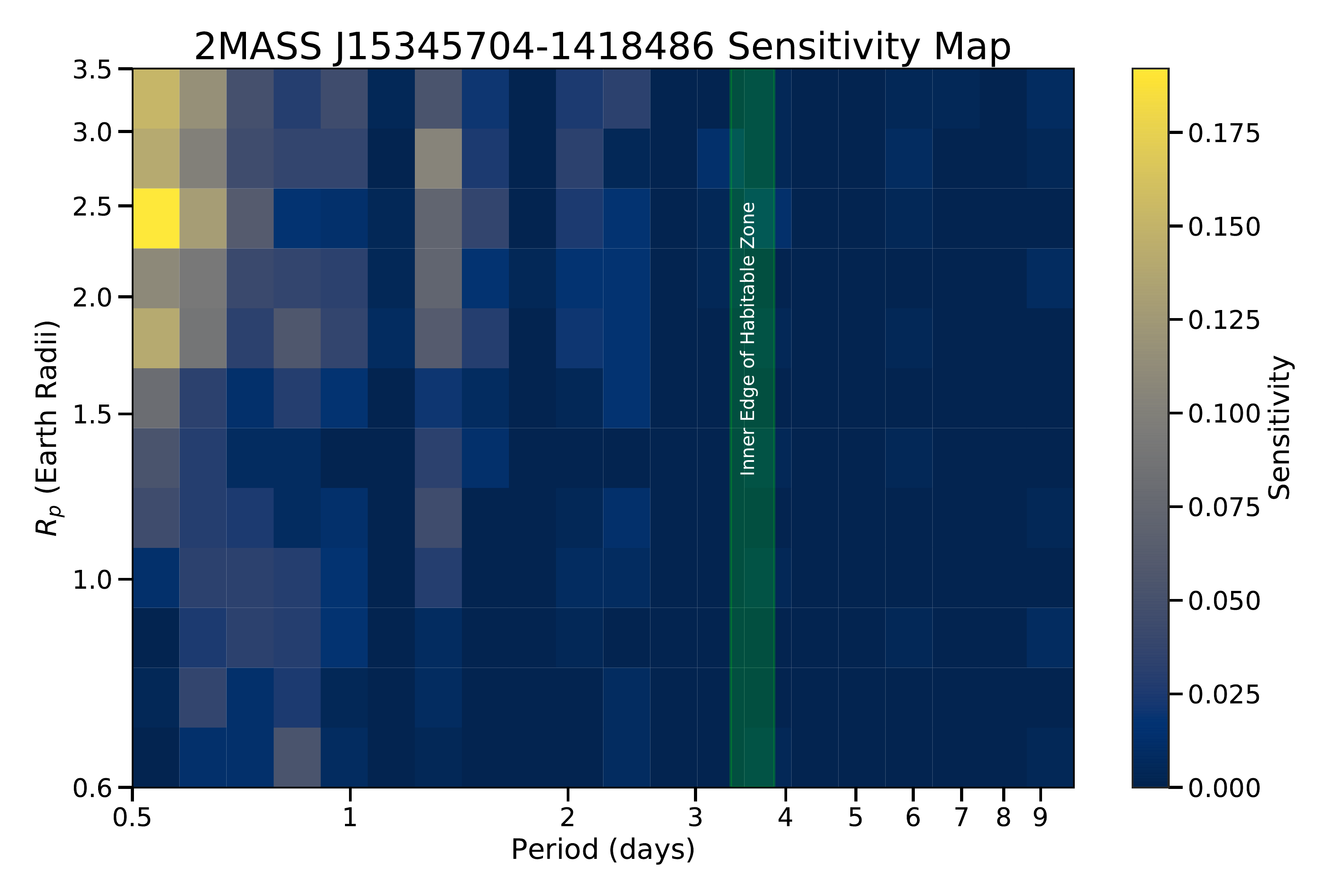}
    \includegraphics[width=0.49\linewidth]{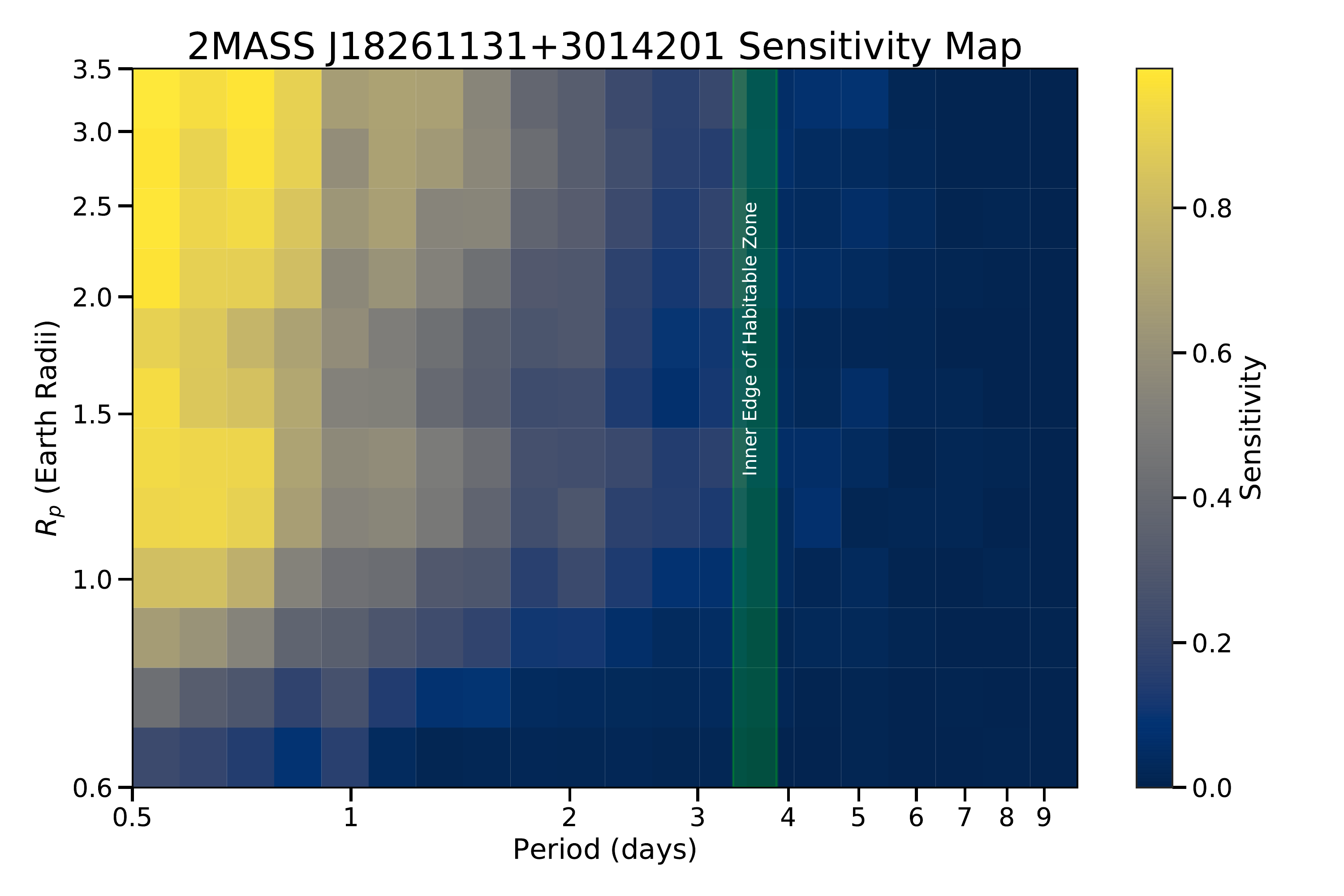}
    \includegraphics[width=0.49\linewidth]{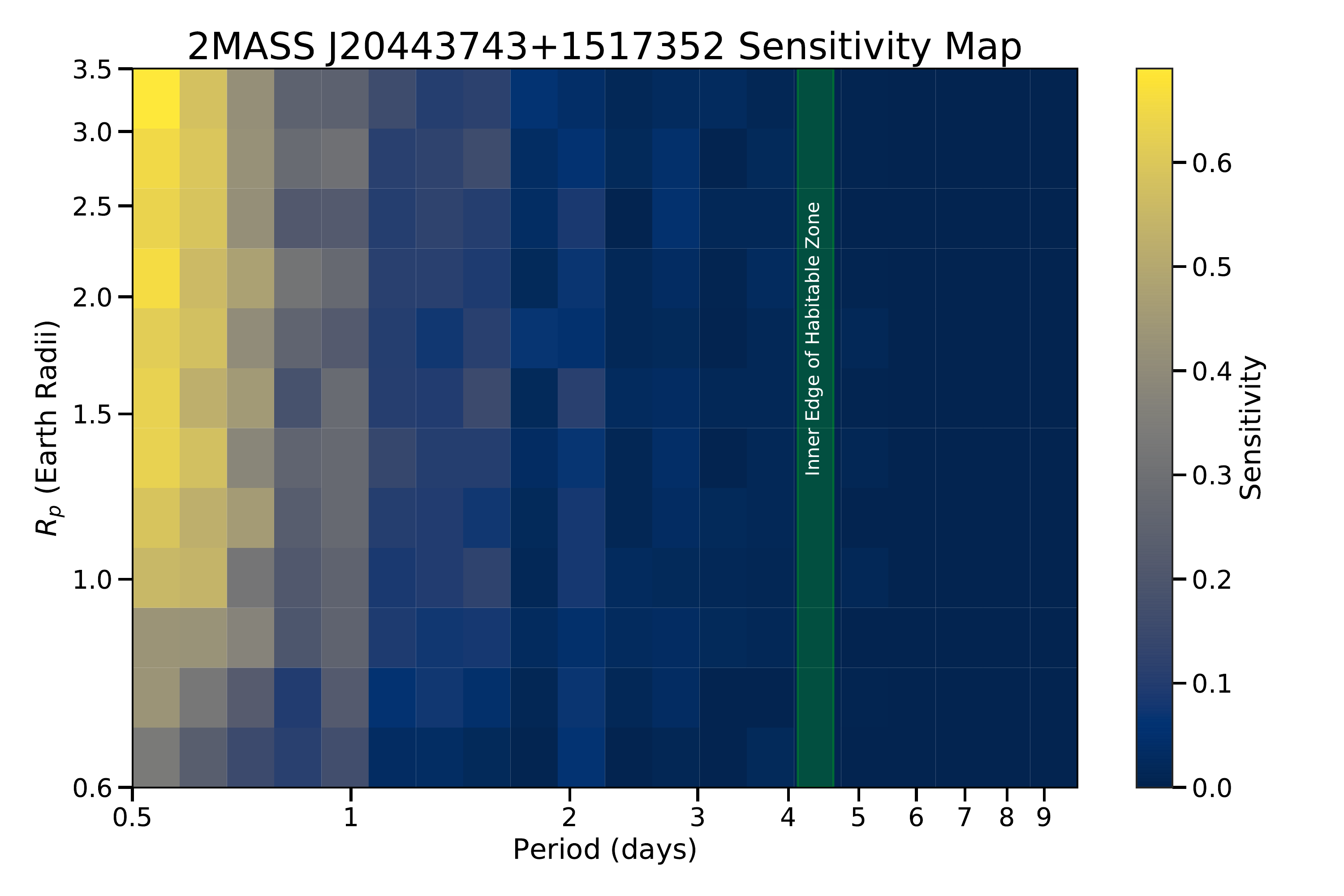}
    \caption{Sensitivity maps for EIC 20, 21, 22, 23}
    \label{fig:sense3}
\end{figure}

\begin{figure}
    \centering
    \includegraphics[width=0.49\linewidth]{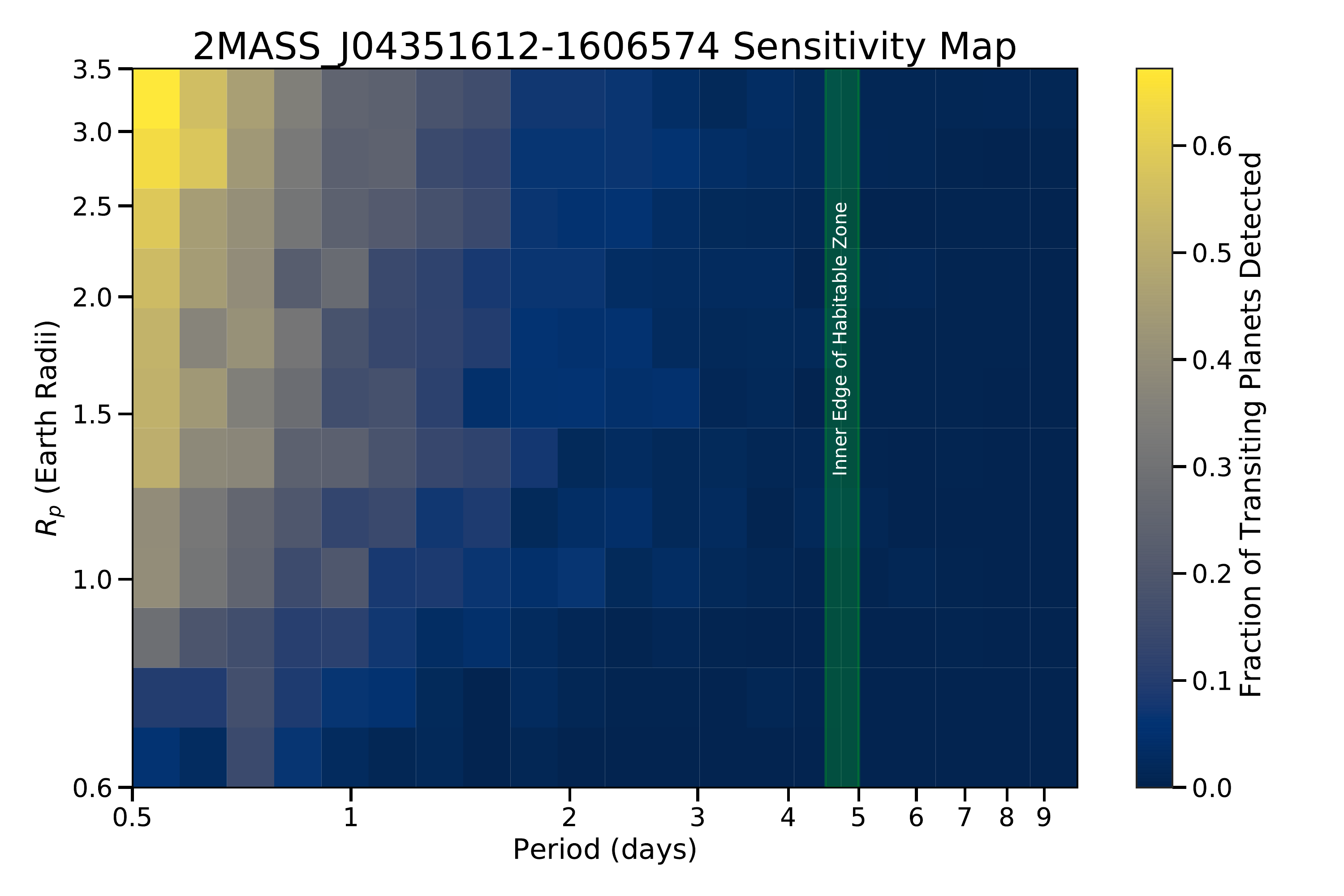}
    \includegraphics[width=0.49\linewidth]{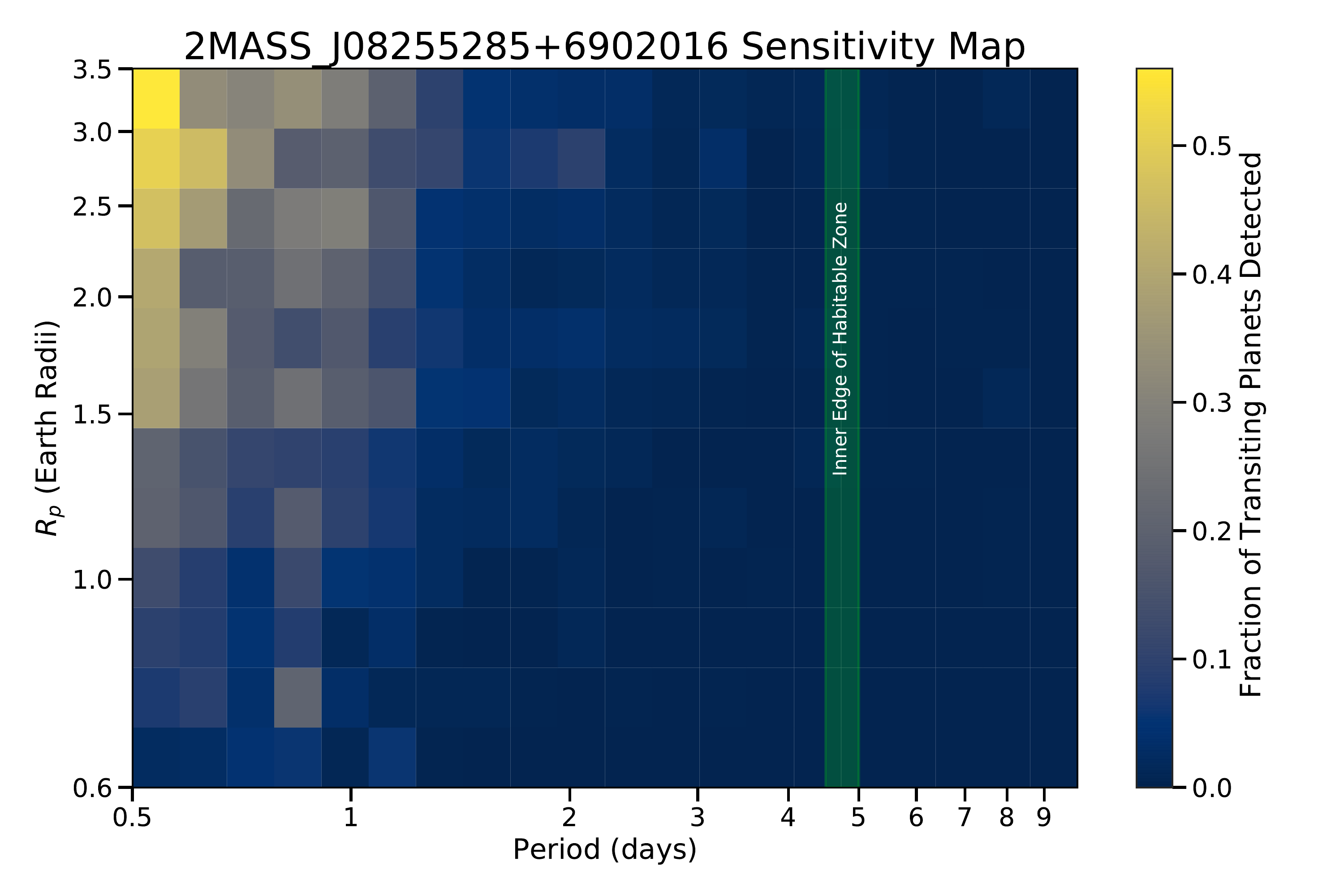}
    \caption{Sensitivity maps for the targets found to be binary stars, EIC 10 and 14}
    \label{fig:sense_binary}
\end{figure}

Here in Figures~\ref{fig:sense1}-\ref{fig:sense3} we show the sensitivity maps for the volume-complete \ntargets{} EDEN survey targets we observed.  The targets that were found to be binaries after the beginning of observations are shown separately in Figure~\ref{fig:sense_binary} at the end.


\section{Other EDEN Observations}

Here we show a table of the other observations taken by Project EDEN on targets at distances of 15-37 pc, earlier spectral types from M2-M6, and for targeted follow-up observations with the TESS Follow-up Observing Program (TFOP).

\begin{table*}[ht]
    \centering
    \caption{EDEN Observations of non-survey and non-follow-up targets}
    \begin{tabular}{cccc}
        \hline
        \textbf{2MASS ID} & \textbf{Distance (pc)} & \textbf{Spectral Type} & \textbf{Hours Observed} \\
        \hline
        \multicolumn{4}{c}{\textit{Observed Targets from 15-21 pc}} \\
        \hline
        J00194579+5213179 & 19.96 & M9 & 314.6\\
        J01242326+6819312 & 20.78 & M7V & 124.3\\
        J01400263+2701505 & 18.99 & M8.5V & 25.3\\
        J03395284+2457273 & 19.27 & M7.7V & 18.1\\
        J05153094+5911184 & 15.22 & M7V & 122.0\\
        J07140394+3702459 & 15.61 & M8 & 294.1\\
        J07410681+1738459 & 15.67 & M7V & 10.0\\
        J08072607+3213101 & 18.88 & M8 & 101.2\\
        J09032096+0540145 & 17.48 & M7V & 111.3\\
        J10163470+2751497 & 17.24 & M8V & 276.5\\
        J10554733+0808427 & 18.75 & M9V & 165.7\\
        J13564148+4342587 & 19.98 & M8V & 189.2\\
        J14032232+3007547 & 19.81 & M9V & 4.5\\
        J15512179+2931062 & 18.48 & M7 & 48.1\\
        J22254853+6421479 & 16.50 & M9.5 & 12.2\\ 
        \hline
        \multicolumn{4}{c}{\textit{Observed non-EDEN Survey Targets}} \\
        \hline
        J00192626+4614078 & 38.34 & M8 & 87.1\\
        J02170993+3526330$^*$ & 10.37 & M7V & 235.6\\
        J02530084+1652532 & 3.83 & dM6 & 5.9\\
        J04351612-1606574$^2$ & 10.604 & M8Ve & 32.9\\
        J05402570+2448090$^*$ & 10.25 & M7V & 51.0\\
        J06083043+4902063 & 37.52 & M5V & 0.7\\
        J07464256+2000321 & 12.36 & L0+L2 & 1.4\\
        J07590587+1523294 & 29.45 & M2V & 3.4\\
        J08022287+0320196 & 27.39 & M4V & 20.9\\
        J08402975+1824091$^*$ & 13.56 & M6V & 6.3\\
        J10022184+4805209 & 14.96 & M1V & 21.0\\
        J10471382+4026493 & 25.73 & M8 & 3.2\\
        J10562886+0700527$^*$ & 2.42 & M6 & 103.8\\
        J11224274+3755484 & 17.69 & M6V & 1.0\\
        J11505787+4822394 & 7.66 & M4.5Ve & 16.4\\
        J12003292+2048513 & 24.62 & M8V & 1.8\\
        J12030930+1701230 & 36.88 & M6V & 1.4\\
        J12555681+5055219 & 21.56 & M4V & 24.4\\
        J13481341+2336486 & 11.88 & M5V & 3.5\\
        J16005083+4019441 & 21.14 & M3 & 0.6\\
        J17151894+0457496 & 14.65 & M4.5V & 68.4\\
        J17351296+2634475 & 15.55 & M7.5+L0 & 91.4\\
        J18393308+2952164 & 12.17 & M6.5Ve & 73.0\\ 
        J18432213+4040209$^*$ & 14.40 & M7.5V & 171.2\\
        J20450403+4429562 & 12.05 & M2V & 15.7\\
        J23415498+4410407 & 3.16 & M5V & 24.9\\
        \hline
    \end{tabular}
    \\[10pt]
    \textbf{Notes}: $^*$: Eruptive or cataclysmic variable star. $^2$ Unresolved binary star EIC 10.
    \label{tab:obsNE}
\end{table*}

\begin{table*}
    \centering
    \caption{EDEN Observations of non-survey targets}
    \begin{tabular}{cccc}
        \hline
        \textbf{TIC ID} & \textbf{Distance (pc)} & \textbf{Spectral Type} & \textbf{Hours Observed} \\
        \hline
        \multicolumn{4}{c}{\textit{Observed TESS Follow-up Observation Program (TFOP) Targets}} \\
        \hline
        8348911 & 51.60 & $\sim$M4V & 3.6\\
        9032367 & 586.30 & $\sim$K3V & 4.5\\
        29918916 & 354.96 & G2IV$^{[1]}$ & 7.8\\
        43064903 & 24.54 & M4.5 & 5.4\\
        66783360 & 254.19 & $\sim$F5.5V & 6.3\\
        98720702 & 99.15 & M3V & 3.2\\
        104208182 & 142.04 & $\sim$K7V& 7.1\\
        142748283 & 48.48 & $\sim$M4V & 4.6\\
        144401492 & 119.24 & K0III & 2.9\\
        175194958 & 190.21 & M1V & 0.6\\
        175241416 & 13.55 & M6V & 4.8\\
        180652891 & 314.98 & Am & 1.7\\
        181804572 & 26.49 & M6V & 25.7\\
        198212955 & 109.52 & $\sim$K6V & 5.1\\
        212957629 & 55.60 & $\sim$M5V & 3.4\\
        232608943 & 216.06 & $\sim$K1V & 6.0\\
        233211762 & 152.00 & $\sim$K7V & 3.8\\
        235678745 & 41.92 & $\sim$M1.5V & 5.3\\
        233602827 & 99.42 & $\sim$M0V & 16.1\\
        237808867 & 967.77 & G0III$^{[1]}$ & 2.4\\
     	240968774 & 37.48 & $\sim$M1V & 1.9\\
     	243185500 & 24.76 & $\sim$M3V & 37.8\\
     	267561446 & 110.60 & $\sim$M0V & 3.2\\
     	269701147 & 53.69 & G0 & 0.9\\
 	 	278892590$^+$ & 12.43 & M7.5e & 11.0\\
 	 	284441182 & 51.74 & $\sim$M1V & 3.5\\
 	 	288185138 & 233.77 & $\sim$F8V & 0.9\\
 	 	357972447 & 256.25 & $\sim$F1V & 1.8\\
        363445338 & 245.53 & K7V & 4.7\\
 	 	368287008 & 47.32 & $\sim$M3.5V & 4.4\\
 	 	377909730 & 440.12 & $\sim$K3V & 4.1\\
 	 	399860444 & 216.67 & G0V & 5.9\\
 	 	408203452$^\dagger$ & N/A & N/A & 12.1\\
 	 	408203470 & 74.80 & $\sim$M3V & 8.2\\
        435931205 & 326.41 & G8IV$^{[1]}$ & 37.1\\
        436584697 & 52.10 & $\sim$M3V & 3.2\\
        439867639 & 48.65 & $\sim$M4V & 2.7\\
        441738827 & 114.93 & $\sim$M1.5V & 6.0\\
        \hline
        \end{tabular}
    \\[10pt]
    \textbf{Notes}: $^\dagger$: Star had zero/negative parallax in Gaia and no stellar parameters in TIC v8.1. $^+$: TRAPPIST-1. Spectral type taken from SIMBAD unless otherwise noted here. $\sim$ indicates value gathered from TIC v8.1 temperature and mass/radius measurements and utilizing the main sequence relations from \citet[][]{Pecaut2013}. [1]: No spectral type in SIMBAD, TIC v8.1 temperature and radius measurements are very similar to stars of these types.
    \label{tab:obsFOP}
\end{table*}


\bibliography{main}{}
\bibliographystyle{aasjournal}



\end{document}